\documentclass{article}
\usepackage{arxiv}
\usepackage[utf8]{inputenc} 
\usepackage[T1]{fontenc}    
\usepackage{hyperref}       
\usepackage{url}            
\usepackage{booktabs}       
\usepackage{amsfonts}       
\usepackage{amsmath}
\usepackage{nicefrac}       
\usepackage{microtype}      
\usepackage{cleveref}       
\usepackage{graphicx}
\usepackage{natbib}
\usepackage{doi}
\usepackage[title]{appendix}

\title{Ensemble reliability and the signal-to-noise paradox in ECMWF subseasonal forecasts}
\author{ \href{https://orcid.org/0000-0002-2958-6637}{\includegraphics[scale=0.06]{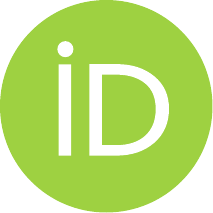}\hspace{1mm}Christopher David Roberts}\\
	ECMWF\\
	Shinfield Park\\
	Reading, United Kingdom\\
	\texttt{chris.roberts@ecmwf.int} \\
	\And
    Frederic Vitart\\
	ECMWF\\
	Shinfield Park\\
	Reading, United Kingdom\\
}
\renewcommand{\headeright}{PREPRINT}
\renewcommand{\undertitle}{PREPRINT}
\renewcommand{\shorttitle}{Ensemble reliability and the signal-to-noise paradox in subseasonal forecasts}

\hypersetup{
  colorlinks   = true, 
  urlcolor     = blue, 
  linkcolor    = blue, 
  citecolor   = red 
}

\begin{document}
\maketitle

\begin{abstract}
Ensemble forecasts sometimes exhibit counterintuitive statistical properties such that the correlation between ensemble means and observations ($r_{mo}$) exceeds the correlation between ensemble means and individual members ($r_{mm}$), and thus models seem to underestimate the predictability of the real world. This behaviour has been interpreted as a `signal-to-noise paradox' (SNP), which is commonly diagnosed using the ratio of predictable components ($\textnormal{RPC} = \sqrt { r_{mo}^2 / r_{mm}^2 }  $). Here, we emphasise the links between ensemble-size-invariant estimates of RPC and other metrics of ensemble reliability and derive a general closed-form expression for RPC in terms of $r_{mo}$, the spread-error ratio (SER), and total variance ratio (VR). Physical constraints on the admissible solutions (i.e. real-valued and non-negative variances) provide a mechanism to identify statistically paradoxical sample estimates of RPC, $r_{mo}$, SER, and VR that correspond to combinations that are not possible without sampling uncertainty. We evaluate three large-scale atmospheric circulation indices in subseasonal reforecasts from the ECMWF Integrated Forecasting System (IFS). Large-ensemble North Atlantic Oscillation (NAO) forecasts evaluated over 80 start dates for the period 2001-2020 generally satisfy unbiased reliability criteria within our estimated sampling uncertainties but still exhibit anomalously high RPC values at some subseasonal lead times. These lead times also coincide with paradoxical combinations of correlation and reliability metrics that are impossible in the large-sample limit, indicating an important role for sampling uncertainties. Nevertheless, wintertime NAO indices averaged over days 16-45 exhibit more robust evidence for unreliability characterised by $\textnormal{RPC}\approx1.5$ suggesting that SNP-like behaviour observed in daily data during the period 2001-2020 is not solely attributable to sampling artefacts. However, these results do not generalise to reduced ensemble size configurations of the same IFS model evaluated over 3120 start dates for the period 1959-2023. In these extended reforecasts, daily NAO indices are well-calibrated and $\textnormal{RPC}\approx1$ for all subseasonal lead times.

\keywords{Subseasonal, seasonal, S2S, predictability, ensemble, reliability, signal, noise, paradox}
\end{abstract}

\section{Introduction}
\label{section:intro}
Ensemble forecast systems are widely used to generate probabilistic weather and climate predictions at lead times of days to decades \citep[e.g. ][]{molteni1996ecmwf, palmer2005representing, doblas2009addressing, vitart2018sub, smith2019robust}. The origins, motivations, and practicalities of ensemble forecasting are comprehensively described by \citet{lewis2005roots} and \citet{leutbecher2008ensemble}. In a statistical sense, the goal of ensemble forecasting is to maximise the sharpness of a predicted distribution subject to reliability \citep{gneiting2007probabilistic}, where reliability indicates statistical consistency between the forecasts and observations. For event-based probabilistic forecasts, reliability requires that the observed frequency of an event tends to $p$ when averaged over many cases for which the event was predicted to occur with probability $p$ \citep{johnson2009reliability, leutbecher2008ensemble, weisheimer2014reliability}. Reliability is also commonly assessed in short- and medium-range ensemble forecasts using a combination of probabilistic verification metrics and comparison of the average ensemble variance with the average squared error of the ensemble mean \citep[e.g.][]{whitaker1998relationship, scherrer2004analysis, hopson2014assessing, yamaguchi2016observation,rodwell2018flow}. 

In contrast, the seasonal-to-decadal forecasting community often emphasises correlation-based evaluation of ensemble mean forecasts, with particular attention given to situations that exhibit the so-called `signal-to-noise paradox' \citep[SNP; ][]{eade2014seasonal, scaife2018signal}. SNP-like behaviour manifests as counterintuitive situations where the correlation between the forecast ensemble mean and the observed truth is larger than the correlation between the forecast ensemble mean and individual forecast members, and thus the real world appears to be more predictable than individual ensemble members from the same forecast model. This type of unreliability has been considered `paradoxical' when it occurs in ensemble forecasts that closely reproduce the total observed variance. An apparent SNP has been identified in a variety of ensemble forecasting systems covering subseasonal to multi-decadal timescales \citep{scaife2014skillful, eade2014seasonal, dunstone2016skilful, scaife2018signal,smith2019robust, garfinkel2024development} and is particularly evident for predictions of the wintertime North Atlantic Oscillation \citep[NAO;][]{baker2018intercomparison}. In particular, \citet{siegert2016bayesian} used a Bayesian framework to evaluate the correlation skill and reliability of seasonal mean winter NAO reforecasts from the Met Office Global Seasonal Forecast System version 5 (GloSea5). They concluded that there was strong evidence (over 99\% certainty) that the GloSea5 reforecasts were not exchangeable with observations due to their underestimation of the magnitude of the predictable component of observed NAO variability. Of particular relevance to the present work is the recent study by \citet{garfinkel2024development}, which diagnoses an apparent SNP in daily mean data from subseasonal reforecasts produced by several models. This study relied on reforecasts with relatively small ensemble sizes and the relevant signal-to-noise diagnostics did not include uncertainty estimates. However, ensemble reliability and signal-to-noise characteristics cannot always be interpreted at face value and should be accompanied by an evaluation of sampling uncertainties \citep[e.g.][]{siegert2016bayesian}.

There is no scientific consensus on the origins or interpretation of the SNP \citep{weisheimer2024signal}. Several studies have proposed physical interpretations of SNP-like behaviour, including deficiencies in the representation of tropical-extratropical teleconnections \citep{scaife2018signal, garfinkel2022winter}, underestimated persistence of non-linear regimes \citep{strommen2019signal, zhang2019understanding}, weak transient eddy feedbacks \citep{scaife2019does, hardiman2022missing}, and inadequate representation of air-sea coupling \citep{zhang2021understanding}. Other studies have emphasised statistical interpretations, including the links to reliability and multi-decadal variations in correlation-based metrics of NAO predictability \citep{shi2015impact, weisheimer2019confident, brocker2023statistical, strommen2023relationship}.

In this study, we emphasise the links between RPC and other metrics of ensemble reliability and evaluate forecast skill, reliability characteristics, and signal-to-noise properties for three large-scale atmospheric circulation indices in subseasonal reforecasts with the European Centre for Medium-Range Weather Forecasts (ECMWF) Integrated Forecasting System (IFS). In particular, we discriminate between sample combinations of correlation and reliability metrics that are statistically paradoxical due to sampling uncertainties and those which provide more robust evidence for unreliability associated with ensemble mean signals that are too weak. There are several novel aspects to our approach, including (i) the combination of evidence from 100-member reforecasts covering 80 start dates for the period 2001-2020 with 10-member extended reforecasts covering 3120 start dates for the period 1959-2023; (ii) our derivation of a general closed-form algebraic expression for RPC in terms of $r_{mo}$, the spread-error ratio (SER), and total variance ratio (VR) and its application to interpreting sampling-related statistical paradoxes; (iii) the careful application of unbiased statistical approaches, including an ensemble-size-invariant estimator for RPC; and (iv) the use of unbiased reliability calibration to distinguish between predictable signals that are too weak and unpredictable noise that is too strong. We combine these reforecast datasets and statistical approaches to answer the following questions: 

\begin{enumerate}
\item \textit{Are ECMWF subseasonal forecasts reliable?}
\item \textit{Do ECMWF subseasonal forecasts exhibit SNP-like behaviour? If yes, in which indices and at what lead times does this behaviour emerge?}
\item \textit{Does reliability calibration provide any insights into the origins of SNP-like behaviour in this dataset?}
\item \textit{Are the answers to the above questions robust to the impacts of sampling uncertainty?}
\end{enumerate}

 The remainder of this paper is organised as follows: Section \ref{section:data} describes the ECMWF subseasonal reforecast datasets and the calculation of large-scale atmospheric circulation indices. Section \ref{section:statistics} provides an overview of the statistical concepts that are relevant for this study. Section \ref{section:results_uncalibrated} evaluates the forecast skill, reliability characteristics, and signal-to-noise properties in uncalibrated daily mean forecasts. Section \ref{section:results_calibrated} evaluates the same, but in forecasts that have been calibrated to enforce reliability. Section \ref{section:monthly_nao} explores the impacts of time-averaging and evaluates three different monthly mean NAO indices separately for summer and winter start dates. Lastly, section \ref{section:conclusions} summarises our results and discusses the challenges of robust and unbiased evaluation of reliability and signal-to-noise properties in the presence of sampling uncertainties.

\section{Data}
\label{section:data}

\subsection{IFS reforecasts}
\label{section:reforecast_data}
We evaluate forecast skill, ensemble reliability, and signal-to-noise properties in subseasonal reforecasts performed with cycle 47r3 of the ECMWF IFS, which includes dynamic representations of the atmosphere, ocean, sea-ice, land-surface, and ocean waves. IFS cycle 47r3 was used operationally at ECMWF from October 12th 2021 to June 27th 2023, when it was replaced by IFS cycle 48r1. \citet{roberts2023euro} provide a more thorough description of IFS cycle 47r3, including an overview of operational and reduced-resolution subseasonal reforecast configurations. Here, we use an experimental reforecast configuration comprised of 46-day, 100-member ensemble forecasts initialised every February 1st, May 1st, August 1st, and November 1st between 2001 and 2020 for a total of 80 start dates. We exclude the unperturbed control forecast (i.e. member 0) from our analysis as it is not statistically exchangeable with perturbed members.  The atmospheric model uses the cubic octahedral reduced Gaussian grid with 137 vertical levels and a horizontal resolution of Tco319 ($\Delta$x$\approx$35 km). The ocean and sea ice models use 75 vertical levels and the eddy-permitting ORCA025 grid ($\Delta$x$\approx$25 km). Otherwise, the IFS configuration, initialization strategy, stochastic parameterizations, and ocean/sea-ice coupling are exactly as described for the operational reforecast configuration used by \citet{roberts2023euro} and will not be repeated here. 

We also present selected results from reduced-resolution reforecasts, which use the same IFS model cycle but configured to use the Tco199 atmospheric grid ($\Delta$x$\approx$50 km) and eddy-parameterized ORCA1 ocean grid ($\Delta$x$\approx$100 km). We use two Tco199 reforecast datasets. The first is a 100-member configuration that covers the same 80 start dates for the period 2001-2020 as the Tco319 reforecasts described above, but at the reduced Tco199 resolution. The second is an extended dataset with a reduced ensemble size of 10 perturbed members, covering 3120 start dates on the 1st, 8th, 15th, and 22nd of each month for the period 1959-2023. These reforecasts represent an extension of the CY47R3\_LR dataset described by \citet{roberts2023euro}. This reduced-resolution Tco199 configuration has previously been shown to be a useful surrogate for higher-resolution configurations with the same IFS cycle \citep{roberts2023euro}. Reforecasts are verified using data from the ERA5 reanalysis \citep{hersbach2020era5}.

\subsection{Atmospheric circulation indices}
\label{section:indices}
We focus our analysis of reliability and signal-to-noise properties on three indices that measure different aspects of the large-scale tropospheric and stratospheric circulation in the Northern Hemisphere. In addition, we evaluate tropical-extratropical teleconnections using lagged composites conditioned on different phases of the Madden-Julian Oscillation (MJO). A brief definition of each index is provided below. 

\subsubsection{The North Atlantic Oscillation (NAO)}
The North Atlantic Oscillation (NAO) is a large-scale mode of atmospheric variability associated with widespread variations in surface weather conditions across Europe and the North Atlantic \citep{hurrell1995decadal}. For each forecast start date, we calculate NAO indices for each forecast member and the equivalent dates in ERA5 by projecting 500 hPa geopotential height anomalies on a regular 2.5$^{\circ}$ $\times$ 2.5$^{\circ}$ latitude-longitude grid onto a precomputed loading pattern.  The NAO loading pattern is defined as the first empirical orthogonal function (EOF) of all-year monthly mean 500 hPa geopotential height anomalies for the period 1979-2018 in the ERA-interim reanalysis \citep{dee2011era} for the region bounded by 20$^{\circ}$N-80$^{\circ}$N and 90$^{\circ}$W-40$^{\circ}$E. EOFs are calculated using the Python `eofs' package \citep{dawson2016eofs} and anomalies are weighted by $\sqrt{\textnormal{cos}(\textnormal{latitude})}$ prior to computation to account for variations in grid-cell area. Forecasts and reanalysis anomalies are projected onto the same observation-based loading pattern and the resulting indices are divided by a precomputed scaling factor, which is defined such that indices can be interpreted as the standardised principal component time-series associated with the EOF-based NAO pattern. In section \ref{section:monthly_nao}, we also evaluate two other definitions of the NAO index: an EOF-based index calculated as above, but using mean sea-level pressure (MSLP) anomalies, and a simple dipole index derived as the difference in MSLP between Lisbon and Reykjavík. 

\subsubsection{The Pacific-North American pattern (PNA)}
The Pacific-North American pattern (PNA) is another large-scale mode of Northern Hemisphere atmospheric variability associated with coherent variations in temperature and precipitation over the North American continent \citep{leathers1991pacific}. We calculate PNA indices following the same procedure outlined above for the NAO. The only difference is that loading patterns are defined from first EOF of monthly mean 500 hPa geopotential height anomalies for the region bounded by 10$^{\circ}$N-80$^{\circ}$N and 150$^{\circ}$E-300$^{\circ}$E. 

\subsubsection{The Northern Hemisphere Stratospheric Polar Vortex (PVORTEX)}
Previous studies have demonstrated that anomalies in the strength of the Northern Hemisphere stratospheric polar vortex can propagate downwards and influence evolution of tropospheric weather regimes such as the NAO \citep{baldwin1999propagation, polvani2004upward, ineson2009role}. We quantify the strength of the Northern Hemisphere stratospheric polar vortex (PVORTEX) in IFS reforecasts and ERA5 as described in \citet{roberts2023euro}, which is consistent with indices used in previous studies to investigate causal links between the troposphere and Northern Hemisphere sudden stratospheric warmings \citep[e.g.][]{limpasuvan2004life, barnes2019tropospheric}. Specifically, indices are calculated from the zonal mean of zonal wind anomalies at 50 hPa and 60$^{\circ}$N and standardised by dividing with a constant factor of 5.15 ms$^{-1}$, which corresponds to the standard deviation of the raw vortex index calculated using all-year daily values from the ERA-interim reanalysis \citep{dee2011era} for the period 1979-2018. 

\subsubsection{The Madden–Julian oscillation (MJO)}
The Madden-Julian Oscillation (MJO) is the leading mode of intraseasonal variability in the tropics \citep{madden1971detection} and an important source of predictability at subseasonal lead times. Variations in tropical convective heating and upper atmosphere circulation anomalies associated with the MJO provide a source of Rossby waves that drive global teleconnections \citep{ hoskins1981steady, sardeshmukh1988generation, cassou2008intraseasonal, lin2009observed}. We diagnose MJO variability using the real-time multivariate MJO (RMM) index following \citet{wheeler2004all} and \citet{gottschalck2010framework}. The two components of the bivariate index (RMM1 and RMM2) are derived by projecting daily mean anomalies onto the two leading observation-based multivariate EOFs of meridionally averaged (15$^{\circ}$S-15$^{\circ}$N) zonal winds at 850 hPa and 200 hPa and outgoing long wave radiation (OLR). MJO amplitude and phase are defined as $\sqrt{\textnormal{RMM1}^2 + \textnormal{RMM2}^2}$ and $\textnormal{arctan2}(\textnormal{RMM2}, \textnormal{RMM1})$, respectively. Phase numbers correspond to the different sectors of MJO phase diagram and are indicative of MJO activity over the Indian Ocean (phases 2 and 3), maritime continent (phases 4 and 5), western Pacific Ocean (phases 6 and 7), and the Atlantic Ocean/Africa (phases 8 and 1).

\section{Statistical concepts}
\label{section:statistics}
To introduce the statistical concepts central to this study, we first consider an idealised \textit{perfectly reliable} ensemble forecast system with $k=1,\dots,N$ members covering $j=1,\dots,M$ independent cases (e.g. forecast start dates). In this idealised system, ensemble forecast members ($x_{1,j},\dots,x_{N,j}$) and the observed truth ($x_{T,j}$) are drawn from the same underlying probability distribution at each start date such that they are statistically exchangeable. 

\subsection{Anomaly calculation}
\label{section:anomalies}
 We define ensemble forecast anomalies ($z_{k,j}$) and observed anomalies ($z_{T,j}$) relative to `by-member--other-years' climatologies following \citet{roberts2025unbiased} such that

\begin{equation}
      z_{k,j} = x_{k,j} - \frac{1}{L-1} \sum\limits^{L}_{\substack{h=1\\h \neq j}} x_{k,h},
\end{equation}
\begin{equation}    
      z_{T,j} = x_{T,j} - \frac{1}{L-1} \sum\limits^{L}_{\substack{h=1\\h \neq j}} x_{T,h},
\end{equation}

where $L$ is the number of years in the reforecast dataset and $h=1,\dots,L$ represents the subset of all cases with the same calendar start date as case $j$. Anomalies are thus calculated relative to climatologies estimated separately for each member and each start date. Crucially, calculating forecast anomalies separately for each member ensures that forecast and verification anomalies are defined relative to reference climatologies with the same sampling uncertainty. This approach has no impact on ensemble means, but ensures that forecast member anomalies remain statistically exchangeable with observed anomalies if the underlying raw forecasts are perfectly reliable. This is not the case for standard approaches to anomaly calculation, which calculate forecast anomalies with respect to a climatology that includes all members. Importantly, this effect is also present for statistics that are not defined in terms of ensemble forecast anomalies but still require the removal of an estimate of the sample mean (e.g. variances, correlations). The statistical justification and motivations for this approach to ensemble forecast anomaly calculation are described in detail by \citet{roberts2025unbiased}. All statistical quantities in this paper are derived from anomalies calculated following the definitions for $z_{k,j}$ and $z_{T,j}$.

\subsection{Ensemble reliability metrics}
\label{section:reliability}
\citet{johnson2009reliability} emphasise that perfectly reliable anomaly-based ensemble forecasts have certain statistical properties, which can be derived from the requirement that observations and forecast members are interchangeable. The first property is that the total variance of the observed truth ($\sigma_T^2 = \mathbb{E} \left[z_{T,j}^2 \right]$) should be equal to the total variance of the ensemble forecast members ($\sigma_z^2 = \mathbb{E} \left[ \left<z_{.,j}^2 \right>_N \right]$) when evaluated over many cases such that

\begin{equation}
      \label{equation:variances}
      \lim_{M\to\infty}  \frac{1}{M}\sum^{M}_{j=1} z_{T,j}^2 = \frac{1}{M}\sum^{M}_{j=1} \left<z_{.,j}^2 \right>_N, 
\end{equation}

where $\mathbb{E} \left[ \cdot \right]$ is the expectation over cases $j$, $\mathbb{E}\left[z_{T,j}\right] = \mathbb{E}\left[z_{k,j}\right] = 0$, and $\left<\cdot\right>_N$ represents the mean over a sample of $N$ members such that the ensemble mean for case $j$ is denoted $\left<z_{.,j}\right>_N \equiv \frac{1}{N}\sum^{N}_{k=1}z_{k,j}$. Following \citet{van2015ensemble} and \citet{roberts2025unbiased}, we refer to this statistical property as \textit{climatological reliability}. 

The second property is that, with appropriate unbiased estimators, the square root of the mean ensemble variance (i.e. `spread') will converge with the root-mean-square error (RMSE) of the ensemble mean such that 

\begin{equation}
      \label{equation:spread_error}
      \lim_{M\to\infty} \left( \frac{\textnormal{Spread}}{\textnormal{RMSE}} \right)_{\textnormal{unbiased}} =  \sqrt {\left( \frac{N+1}{N-1} \right) } \frac{\sqrt{\frac{1}{M}\sum^{M}_{j=1} \left< \left( z_{.,j} - \left<z_{.,j}\right>_N\right)^2\right>_N }}{\sqrt{\frac{1}{M}\sum^{M}_{j=1}\left( z_{T,j} - \left<z_{.,j}\right>_N \right)^2 }} = 1,
\end{equation} 
where the factor of $\sqrt{\frac{N+1}{N-1}}$ ensures estimates are unbiased with ensemble size as discussed by \citet{leutbecher2008ensemble}. We refer to this spread-error relationship as \textit{ensemble variance reliability}.

We consider these metrics to be measures of \emph{unconditional} reliability as they measure statistical consistency in expectation across all cases. This does not necessarily imply \emph{conditional} reliability, i.e. statistical consistency within arbitrary subsets of forecasts (e.g. grouped by spread or forecast regime). In other words, these unconditional reliability criteria are necessary but not sufficient conditions for a perfectly reliable ensemble forecast (i.e. members and observations drawn from the same underlying distribution at each start date).


\subsection{Correlations}
\label{section:correlations}
\citet{johnson2009reliability} also highlighted the links between reliability and correlation-based evaluation of ensemble mean forecasts by considering the impact of a simple member-by-member statistical calibration that enforces ensemble reliability. They showed that, in the limit\footnote{This limit is not mentioned by \citet{johnson2009reliability}, but it can be inferred from equations \ref{equation:variances} and \ref{equation:spread_error}.} $M,N\rightarrow\infty$, a calibration that simultaneously enforces climatological reliability (equation \ref{equation:variances}) and ensemble variance reliability (equation \ref{equation:spread_error}) is exactly equivalent to a calibration that enforces equation \ref{equation:variances} combined with the constraint that the correlation between the forecast ensemble mean and observations ($r_{mo}$) is equal to the correlation between forecast ensemble mean and forecast members ($r_{mm}$). For a finite ensemble size, the relevant correlations can be defined as follows

\begin{equation}
      \label{equation:correlation}
      r_{mo}= \frac{\mathbb{E}\left[ \left<z_{.,j}\right>_{N-1}  z_{T,j}  \right]}{\sqrt{\mathbb{E}\left[ \left<z_{.,j}\right>_{N-1} ^2 \right]   \mathbb{E}\left[ z_{T,j}^2  \right]}},
\end{equation}

\begin{equation}
      \label{equation:model_member_correlation}
      r_{mm(k)} = \frac{\mathbb{E}\left[ \left<z_{.,j}\right>_{N-1}^{i \ne k}   z_{k,j}  \right]}{\sqrt{\mathbb{E}\left[ \left( \left<z_{.,j}\right>_{N-1}^{i \ne k}  \right)^2 \right]   \mathbb{E}\left[ z_{k,j}^2  \right]}},     
\end{equation}

\begin{equation}
      \label{equation:model_mean_correlation}
      \overline{r_{mm}} = \frac{1}{N} \sum\limits^{N}_{k=1} r_{mm(k)} ,
\end{equation}
where we define $\left<\cdot\right>_{N-1}$ to indicate the ensemble mean constructed from the first $N-1$ members and $\left<\cdot\right>_{N-1}^{i \ne k} \equiv \frac{1}{N-1}\sum\limits^{N}_{\substack{i=1\\i \ne k}}$ such that $\left<z_{.,j}\right>_{N-1}^{i \ne k}$ represents the ensemble mean for case $j$ after excluding member $k$. The value of $r_{mm(k)}$ thus represents the `model-model' correlation between the forecast ensemble mean and an excluded ensemble member and $\overline{r_{mm}}$ represents the mean of $N$ estimates of $r_{mm(k)}$. We use this definition of $r_{mm(k)}$ for consistency with $r_{mo}$, for which the forecast ensemble means do not include the observed value. Importantly, we also calculate $r_{mo}$ using an ensemble mean constructed from $N-1$ members for consistency with $r_{mm(k)}$. The use of $N-1$ rather than $N$ members ensures that estimates of $r_{mo}$ are exchangeable with estimates of $r_{mm(k)}$ in a perfectly reliable ensemble.

However, equations \ref{equation:model_member_correlation} and \ref{equation:model_mean_correlation} are not the only way to estimate $r_{mm(k)}$ and $\overline{r_{mm}}$. In a well-constructed ensemble, the members for case $j$ can be considered independent draws from the same underlying probability distribution and there is no particular reason that $r_{mm(k)}$ should be estimated using the same excluded member $k$ for each case $j$. For example, we also estimate model-model correlations using $r_{mm(\mathbf{krandom})}$, where $\mathbf{krandom} = (k_j)_{j=1}^{M}$ represents a vector of excluded members that are fixed over forecast lead times but selected randomly for each start date $j$. The $N$ estimates of $r_{mm(k)}$ are thus a subset of the $N^M$ possible estimates of $r_{mm(\mathbf{krandom})}$. Given that estimates of $r_{mm(k)}$ and $r_{mm(\mathbf{krandom})}$ are statistically exchangeable, we refer to both methods for calculating model-model correlations using the notation $r_{mm(k)}$ and provide clarification on the sampling methods in the associated text or figure captions.

\subsection{The ratio of predictable components}
\label{section:rpc}
As described in section \ref{section:intro}, the relationship between sample estimates of $r_{mo}$ and $\overline{r_{mm}}$ in ensemble forecasting systems has drawn significant attention in the climate forecasting community in the context of the SNP \citep{eade2014seasonal, scaife2018signal}. The SNP was originally diagnosed using a variance-based definition of the ratio of predictable components \cite[RPC; ][]{eade2014seasonal} defined in terms of $r_{mo}$, $\sigma_z^2$, and $\sigma_{\left<z\right>}^2=\mathbb{E}\left[ \left<z_{.,j}\right>_{N} ^2 \right]$ as

\begin{equation}
\label{equation:rpc_var}
      \textnormal{RPC}_{\textnormal{Var}} = \frac{r_{mo}}{\sqrt{\sigma_{\left<z\right>}^2 / \sigma_z^2}},
 \end{equation}
which is biased low for finite $N$, though \citet{hardiman2022missing} have proposed an alternative form that is less sensitive to ensemble size. An alternative expression for RPC can be defined directly from correlations \citep{scaife2018signal} as 

\begin{equation}
\label{equation:rpc_corr}
      \textnormal{RPC} = \sqrt {\frac{r_{mo}^2}{\overline{r_{mm}}^2}}.
 \end{equation}
The only requirement for this correlation-based definition of RPC to be unbiased with ensemble size is that estimates of $r_{mo}$ and $r_{mm(k)}$ are exchangeable when the underlying forecast anomalies are also exchangeable (i.e. when the forecast is reliable). This condition is satisfied when $r_{mo}$  and $r_{mm(k)}$ are calculated using equations \ref{equation:correlation} and \ref{equation:model_member_correlation}, respectively (see figure \ref{fig:biased_vs_unbiased_rpc}). In all forms, RPC values that significantly exceed one are evidence of unreliability and have been interpreted as manifestations of a predictability paradox. To accompany correlation-based estimates of RPC, we also calculate empirical distributions of RPC$_{mm(k)}$, which represent the model-model equivalents of RPC calculated following our notation for excluded members described in section \ref{section:correlations}. 

To emphasize the links between the RPC and other metrics of reliability, we also derive a general closed-form algebraic expression for RPC in terms of the population correlation between the forecast ensemble mean and observations ($\rho_{mo}$), the spread-error ratio (SER; equation \ref{equation:spread_error}), and total variance ratio ($\textnormal{VR} = \sigma_z^2 / \sigma_T^2$)

\begin{equation}
\label{equation:rpc_solutions}
\textnormal{RPC}_{1,2} = \frac{\sqrt{\textnormal{VR}} \left( 1 + \textnormal{SER}^2 \right)}{ \textnormal{SER}^2  \pm  \rho_{mo}^{-1}  \sqrt{\Delta}},
\end{equation}
where $\textnormal{RPC}_{1,2}$ represents two solution families corresponding to the positive and negative roots of a quadratic equation for $\sigma_{\left<z\right>}$ and $\Delta$ is a function of SER, VR, and $\rho_{mo}$. Additional details are provided in appendix A, which includes a full derivation of equation \ref{equation:rpc_solutions}, the definition of $\Delta$, and the conditions for physically admissible solutions (i.e. variances must be real-valued and non-negative). The solutions to equation \ref{equation:rpc_solutions} require no distributional assumptions or conditions on the statistical exchangeability of forecasts and observations and hold in the large-sample limit, where population quantities are well-defined and sampling uncertainty disappears.

\begin{figure}[!htbp]
    \centering           
    \includegraphics[width=15cm]{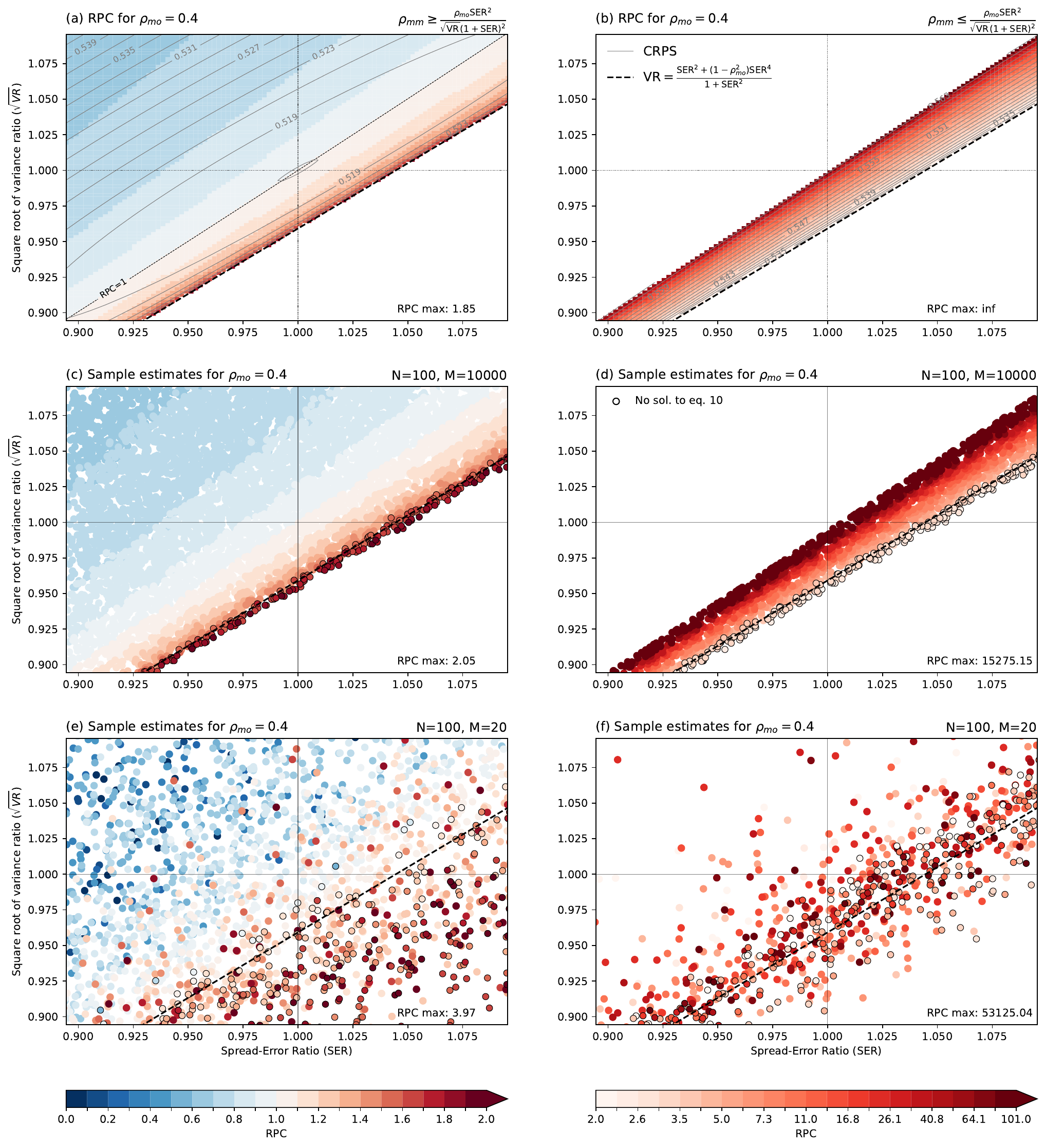}
    \caption{(a, b) RPC solutions calculated using equation \ref{equation:rpc_solutions} with $\rho_{mo}=0.4$ and specified values for spread-error ratio (SER) and total variance ratio (VR). Masked values correspond to regions of SER-VR space without physical solutions to equation \ref{equation:rpc_solutions} as described in appendix A. Grey contours correspond to the expected value of the continuous ranked probability score (CRPS) calculated for $\sigma_T = 1.0$ following \citet{leutbecher2021understanding}. (c, d) Sample estimates of RPC derived using equation \ref{equation:rpc_corr} using $M=10000$ independent cases and $N=100$ members of idealised ensemble forecast and observational data generated from multivariate Gaussian distributions with population parameters that satisfy $\rho_{mo}=0.4$ and the values of SER and VR plotted in panels (a) and (b). Black circles indicate sample combinations of $r_{mo}$, SER, and VR that have no solution to equation \ref{equation:rpc_solutions}. Note that these solutions do not exactly correspond with the masked regions in panels (a) and (b) as the physical admissibility of each data point is evaluated independently using the sample estimate of $r_{mo}$ rather than the population value of $\rho_{mo}$. (e, f) As panels (c) and (d) but calculated using only $M=20$ independent cases.}
    \label{fig:rpc_vs_ser_vr}
\end{figure}

Figures \ref{fig:rpc_vs_ser_vr}a-b illustrate the two RPC solution families in SER-VR space for $\rho_{mo}=0.4$. The first solution family (figure \ref{fig:rpc_vs_ser_vr}a) corresponds to population model-model correlations ($\rho_{mm}$) that exceed a threshold value and includes the solution that minimizes the continuous ranked probability score (CRPS) and satisfies climatological and ensemble reliability such that $\text{RPC} \to 1$ as $\text{SER},\text{VR} \to 1$. Figure \ref{fig:rpc_vs_ser_vr}a also clearly illustrates that RPC alone provides an incomplete description of the reliability characteristics of a forecast system such that the constraint $\textnormal{RPC}=1$ corresponds to a continuum of solutions forming a line in SER-VR space. Furthermore, small deviations from reliability towards over-dispersive (i.e. $\text{SER} > 1$) or under-active (i.e. $\text{VR} < 1$) solutions can result in substantial increases in RPC. This result is consistent with previous work, which has highlighted that `approximately reliable' forecasts can still exhibit anomalously high RPC values \citep{brocker2023statistical}. The sensitivity of RPC to changes in SER and VR is particularly pronounced when correlations are lower, as shown in supplementary figure \ref{fig:rpc_vs_ser_vr_rho02}. More generally, the extent of the physically inadmissible region in SER-VR space depends strongly on the correlation, as illustrated in supplementary figures \ref{fig:rpc_vs_ser_vr_rho02} and \ref{fig:rpc_vs_ser_vr_rho06}.

The second solution family exhibits divergent behaviour, such that $\rho_{mm} \to 0$ and $\text{RPC} \to \infty$ as $\text{SER},\text{VR} \to 1$ (figure \ref{fig:rpc_vs_ser_vr}b). This limiting behaviour represents the trivial solution in which the climatological and ensemble variance reliability conditions are formally satisfied, but the ensemble-mean variance collapses to zero and CRPS is increased. This branch of solutions is characterised by large and positive values of the RPC (figure \ref{fig:rpc_vs_ser_vr}b) associated with a weak ensemble mean signal.

\subsection{Ensemble calibration}
\label{section:calibration}
In section \ref{section:results_calibrated} we use an unbiased member-by-member calibration approach that simultaneously enforces climatological reliability (equation \ref{equation:variances}) and ensemble variance reliability (equation \ref{equation:spread_error}). This calibration ensures that forecast anomalies satisfy equations \ref{equation:variances} and \ref{equation:spread_error}, which are properties of a perfectly reliable ensemble, when averaged over a sample of start dates.  Calibrated forecast anomalies ($\hat{z}_{k,j}$) are derived by separately modifying the ensemble mean and perturbations from the ensemble mean as follows

\begin{equation}
    \label{equation:calibration}
    \hat{z}_{k,j} = \alpha \left<z_{.,j}\right>_N + \beta \left( z_{k,j} - \left<z_{.,j}\right>_N   \right),
\end{equation}

where  

\begin{equation}
      \label{equation:alpha}
     \alpha = \frac{\sigma_{T}}{\sigma_{\left<z\right>}} \left( \frac{r_{mo} + \sqrt{r_{mo}^2 + R^2 - 1}}{R+1} \right),
  \end{equation}

  \begin{align}
      \label{equation:beta}
      \beta^2  = \frac{\sigma_{T}^2 - \alpha^2 \sigma_{\left<z\right>}^2}{\mathbb{E}[(z_{k,j} - \left<z_{.,j}\right>_N)^2 ]},
  \end{align}

and $R = \frac{N+1}{N-1}$. This formulation follows \citet{johnson2009reliability} and has a long history in seasonal forecasting \citep[e.g.][]{von1999use, doblas2005rationale}. The novelty of our approach is to estimate parameters $\alpha$ and $\beta$ following \citet{roberts2025unbiased} such that they are unbiased with ensemble size resulting in adjusted ensemble forecasts that exactly satisfy the climatological reliability and unbiased ensemble variance reliability conditions described in section \ref{section:reliability}, even for small ensemble sizes. 

Importantly, in the limit $R\to 1$, this member-by-member reliability calibration is algebraically identical to regression-based approaches to correct for the signal-to-noise paradox \citep[e.g. ][]{eade2014seasonal}. As shown in figure \ref{fig:rpc_vs_ser_vr}a, RPC $\rightarrow$ 1 as $M,N\rightarrow \infty$ when forecasts satisfy the climatological reliability and ensemble variance reliability criteria described by equations \ref{equation:variances} and \ref{equation:spread_error}. This equivalence was previously described by \citet{johnson2009reliability}, where they demonstrated that estimates of $\alpha$ derived by enforcing either reliability or correlation-based constraints are identical to the linear regression slope coefficient that minimizes the mean squared error between the adjusted ensemble mean and the observations. 

Our estimate of $\alpha$ can thus be interpreted as an unbiased estimate of the regression-based correction for the SNP that would be achieved with an infinite ensemble size \citep[e.g. ][]{eade2014seasonal} and is equivalent to the inverse of the coupling parameter that determines the sensitivity of forecasts to the predictable signal in the signal-plus-noise model of \citet{siegert2016bayesian}. The correction for the signal-to-noise paradox can thus be interpreted as a reliability calibration and an apparent SNP can occur because the predictable signal is too weak (i.e. the diagnosed value of $\alpha > 1$) and/or the unpredictable noise is too large (i.e. the diagnosed value of $\beta < 1$). Furthermore, the coupling of VR, SER, and RPC through linear reliability calibration implies that RPC is also a metric of unconditional reliability. In particular, this calibration enforces $\textnormal{VR} = \textnormal{SER} = \textnormal{RPC} = 1$ when averaged over all start dates, but does not require or imply conditional reliability. In other words, it does not require statistical consistency within subsets of forecasts grouped by, for example, the state of a physical driver such as the MJO.

\subsection{Sampling uncertainty}
\label{section:uncertainty}
\subsubsection{Bootstrap resampling methods}
For a perfectly reliable ensemble forecast system, sample estimates of $\overline{r_{mm}}$, $r_{mm(k)}$, and $r_{mo}$ will converge with the underlying population correlation, $\rho$, when evaluated using a sufficiently large ensemble size, $N$, over a sufficiently large number of independent cases, $M$. We thus expect RPC $\rightarrow$ 1 as $M,N \rightarrow \infty$ in a well-calibrated model. However, apparent unreliability and SNP-like behaviour will still sometimes occur in a perfectly reliable ensemble forecast system as a natural consequence of sampling uncertainty when $M$ and $N$ are finite. 

\begin{figure}[!htbp]
      \centering           
      \includegraphics[height=18cm]{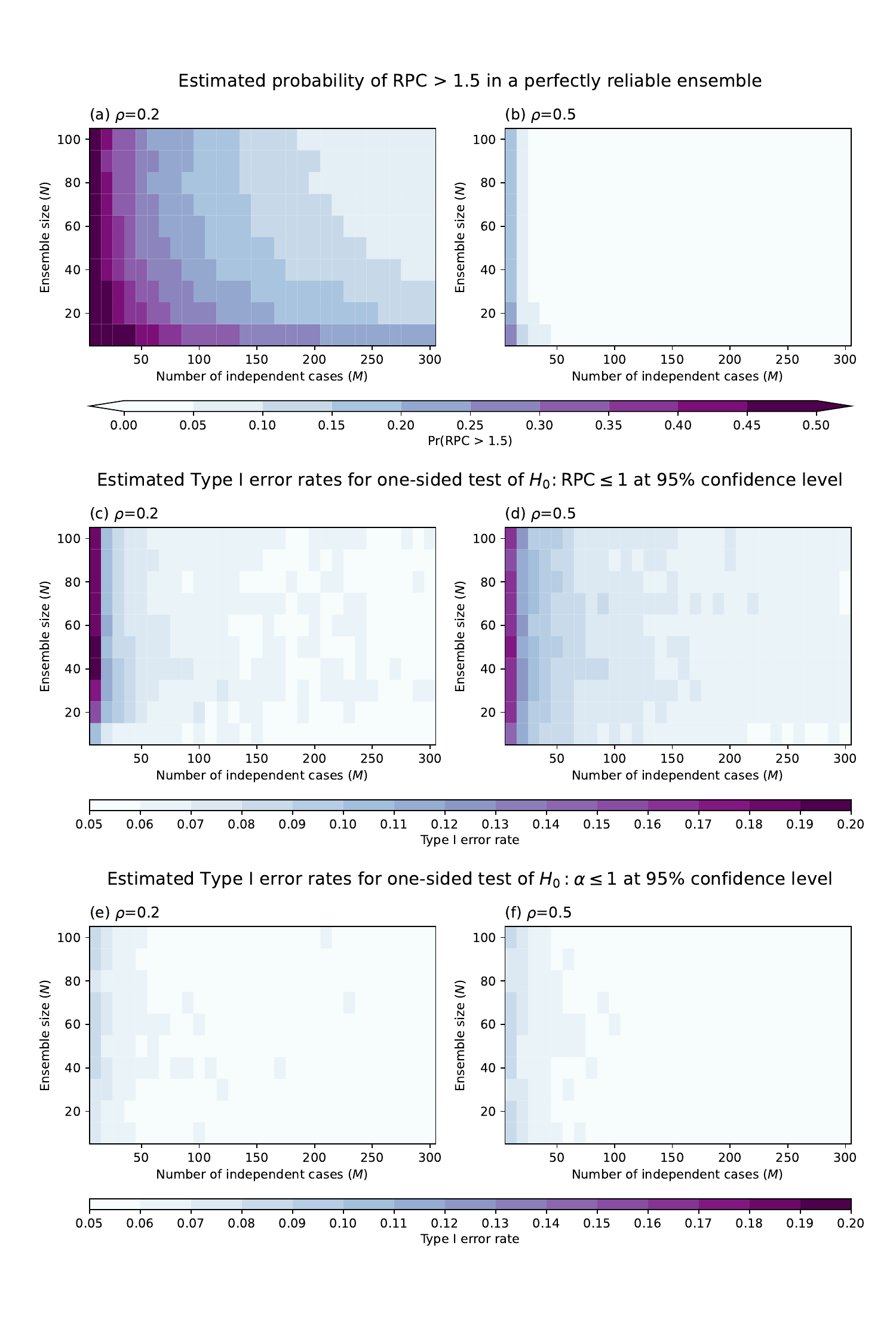}
      \caption{(a-b) Estimated probability that RPC exceeds 1.5 in idealised perfectly reliable ensembles as function of the number of independent cases ($M$), ensemble size ($N$), and different levels of intrinsic correlation skill ($\rho$). Each estimate of Pr(RPC$ > 1.5$) is derived from a distribution of 10,000 RPC values generated from an idealised perfectly reliable ensemble dataset. Forecast and observations are generated by the same process such that $z_{k,j} = \left<z_{.,j}\right>_N + n_{k,j}$, where $\left<z_{.,j}\right>_N \sim \mathcal{N}(0,\sigma_{\left<z\right>}^2)$ is a predictable component common to all members and observations and $n_{k,j} \sim \mathcal{N}(0,\sigma_{\epsilon}^2)$ is an unpredictable noise component such that $\sigma_{z}^2 = \sigma_{\left<z\right>}^2 + \sigma_{\epsilon}^2$. (c-d) Estimated type I error rates for the null hypothesis that RPC $\leq 1$ based on 95\% confidence intervals derived by bootstrap resampling (with replacement) applied to the idealised data used in panels (a) and (b). (e-f) Estimated type I error rates for the null hypothesis that $\alpha \leq 1$ based on 95\% confidence intervals derived by bootstrap resampling (with replacement) applied to the idealised data used in panels (a) and (b).}
      \label{fig:rpc_vs_M_and_N}
  \end{figure}

To illustrate this point, figure \ref{fig:rpc_vs_M_and_N}a-b shows the probability of RPC exceeding a threshold value of 1.5 as a function of $M$ and $N$ in an idealised perfectly reliable ensemble. When intrinsic predictability is low (i.e. $\rho = 0.2$), there is a 30-35\% chance of RPC exceeding 1.5 for $N=100$ and $M=30$, even when forecasts and observations are generated by the same statistical process. This is reduced to $\sim$5\% if RPC is evaluated using $N=100$ and $M=300$. Importantly, the definition of RPC means that these empirical distributions are not symmetric around RPC=1 for low predictability and small sample sizes. For example, with $\rho = 0.2$ there is just a 20-25\% chance of RPC less than 0.5 for $N=100$ and $M=30$ (not shown). If intrinsic predictability is modest (i.e. $\rho = 0.5$), the probability of detecting RPC$ > 1.5$ is dramatically reduced (figure \ref{fig:rpc_vs_M_and_N}b).  If intrinsic predictability is high (i.e. $\rho > 0.7$) and $M$ and $N$ are sufficiently large such that $\overline{r_{mm}} \to \rho$, then RPC$ > 1.5$ becomes impossible.  


For this reason, it is important that point estimates of RPC and other metrics of forecast reliability are accompanied by reliable confidence intervals\footnote{A 95\% confidence interval for a parameter estimate is considered reliable if it contains the true parameter 95\% of the time across many independent samples.} to assess statistical significance. \citet{siegert2016bayesian} proposed a Bayesian framework for evaluation of ensemble forecasts that provides robust uncertainty estimates for sample statistics (e.g. correlations and signal-to-noise ratios) and the parameters of a statistical model describing the joint distribution of forecast members and observations. However, Bayesian methods can be computationally expensive, and the specification of suitable prior distributions can require expert judgement when uninformative priors are inadequate \citep{siegert2016bayesian}. For these reasons, it is not trivial to generalise such Bayesian approaches to ensemble reforecast data covering a range of variables, regions (i.e. indices or grid points), and lead times.

We follow previous studies \citep[e.g.][]{eade2014seasonal, roberts2023euro} and estimate uncertainties in forecast reliability and signal-to-noise metrics using empirical distributions derived by bootstrap resampling (with replacement) from the available forecast start dates \citep[e.g. ][]{efron1994introduction, wilks2011statistical}. Statistically robust SNP-like behaviour associated with a weak predictable signal is diagnosed when RPC$ > 1$, $\alpha > 1$, and their associated confidence intervals do not overlap with one. However, an important caveat to this approach is that the resulting confidence intervals for RPC are not generally reliable for small sample sizes (figure \ref{fig:rpc_vs_M_and_N}c-f). 

For example, 95\% confidence intervals for the null hypothesis that $\mathrm{RPC} \leq 1$ derived from perfectly reliable model data can have Type I error rates exceeding the nominal 0.05 level (figure \ref{fig:rpc_vs_M_and_N}c-d), which could result in overconfident diagnosis of forecast unreliability. These inflated Type I error rates are most pronounced for small samples (i.e. $M < 50$) and occur for two reasons. Firstly, for small sample sizes and/or low predictability situations, empirical distributions of RPC derived by bootstrap resampling are positively skewed due to the impact of very small and/or negative sample correlations. Secondly, bootstrap resampling approaches to the estimation of confidence intervals are known to exhibit inflated Type I error rates when applied to small sample sizes that are not representative of the full distribution \citep{diciccio1987bootstrap, koopman2015small}. Furthermore, this effect appears to be amplified at larger ensemble sizes because of the asymmetry in the sampling characteristics of $r_{mo}$ (a single estimate for each bootstrap resample) and $\overline{r_{mm}}$ (an average of $N$ leave-one-out estimates), which can lead to bootstrap confidence intervals for RPC that are even narrower when $M$ is small and $N$ is large. This effect is less pronounced for the calibration parameter $\alpha$ (figure \ref{fig:rpc_vs_M_and_N}e-f), though Type I errors remain slightly inflated for smaller sample sizes. 

Given these potential issues with our bootstrap estimates of sampling uncertainty, we also directly compare estimates of $r_{mo}$ and RPC with empirical distributions of their model-model equivalents, which are derived by either systematically or randomly excluding a single member as the `truth' for each start date as described in sections \ref{section:correlations} and \ref{section:rpc}. In this case, forecast unreliability and SNP-like behaviour are identified when $r_{mo}$ and RPC do not plausibly lie within the empirical distributions of model-model equivalents. Specifically, we calculate the percentage of model-model estimates of RPC$_{mm(k)}$ that exceed the sample estimate of RPC, which can be interpreted as an empirical one-sided p-value for the null hypothesis that the forecast system is perfectly reliable. The benefit of this approach is that it does not make any assumptions about the expected value of finite sample statistics from a perfectly reliable model as biases will be common to both model-observation estimates and the empirical model-model distribution.

\subsubsection{Statistical paradoxes and sampling uncertainty}
VR, SER, and RPC measure different aspects of statistical consistency, but they are coupled through linear reliability calibration and must be mutually consistent. Equation \ref{equation:rpc_solutions} directly links all three metrics in the large-sample limit and thus provides a theoretical framework to collectively evaluate estimates of $r_{mo}$, VR, SER, and RPC without making any distributional assumptions. This allows the identification of seemingly paradoxical finite sample combinations of SER, VR, and RPC in the sense that they are qualitatively different from those possible using equation \ref{equation:rpc_solutions}. Furthermore, physical constraints on the RPC solution space (i.e. real-valued and non-negative variances) mean that some sample combinations of $r_{mo}$, SER, and VR are not possible in the large-sample limit (i.e. $M,N \to \infty$) and thus cannot be interpreted together as valid estimates of the converged population quantities. In other words, sample estimates of reliability and correlation metrics without solutions to equation \ref{equation:rpc_solutions} can be considered statistically paradoxical in the sense that they represent a combination that is impossible without the influence of sampling uncertainty. 

The likelihood of identifying sample estimates of SER, VR, and $r_{mo}$ without solutions to equation \ref{equation:rpc_solutions} depends on several factors, including the number of independent forecast start dates ($M$), ensemble size ($N$), and the (unknown) true forecast and observation distributions. The absence of solutions to equation \ref{equation:rpc_solutions} can occur for both reliable and unreliable forecasts and provides a mechanism to identify situations where sampling uncertainties are sufficient to preclude a naive interpretation of sample estimates of SER, VR, and RPC as values that would be achieved in the limit $M,N \to \infty$.

Figures \ref{fig:rpc_vs_ser_vr}c-f illustrate sample estimates of RPC derived using equation \ref{equation:rpc_corr} for idealised ensemble forecast and observational data generated using a multivariate Gaussian distribution with population parameters that satisfy the values of $\rho_{mo}$, SER, and VR shown in figures \ref{fig:rpc_vs_ser_vr}a-b. Idealised data derived using $N=100$ members and $M=10000$ independent cases yields sample RPC estimates that are clustered closely around the solutions that would be achieved using equation \ref{equation:rpc_solutions} and the underlying population parameters (figures \ref{fig:rpc_vs_ser_vr}c-d). However, it is still possible to identify some estimates that correspond to sample combinations of $r_{mo}$, SER, and VR without solutions to equation \ref{equation:rpc_solutions}. In contrast, idealised observational and ensemble forecast data derived using $N=100$ and $M=20$ show much large sampling variability, with many estimates of RPC $> 1$ lying outside the admissible solution space. It is also possible to identify sample estimates of RPC that are qualitatively different from those possible using equation \ref{equation:rpc_solutions} (e.g. sample estimates with RPC $>1$ despite sample estimates of VR$ > 1$ and SER$ < 1$). We use these theoretical constraints in combination with bootstrap uncertainty estimates to interpret the reliability characteristics of IFS ensemble forecasts. 


\section{Uncalibrated daily mean forecasts}
\label{section:results_uncalibrated}
\begin{figure}[!htbp]
      \centering           
      \includegraphics[width=15cm]{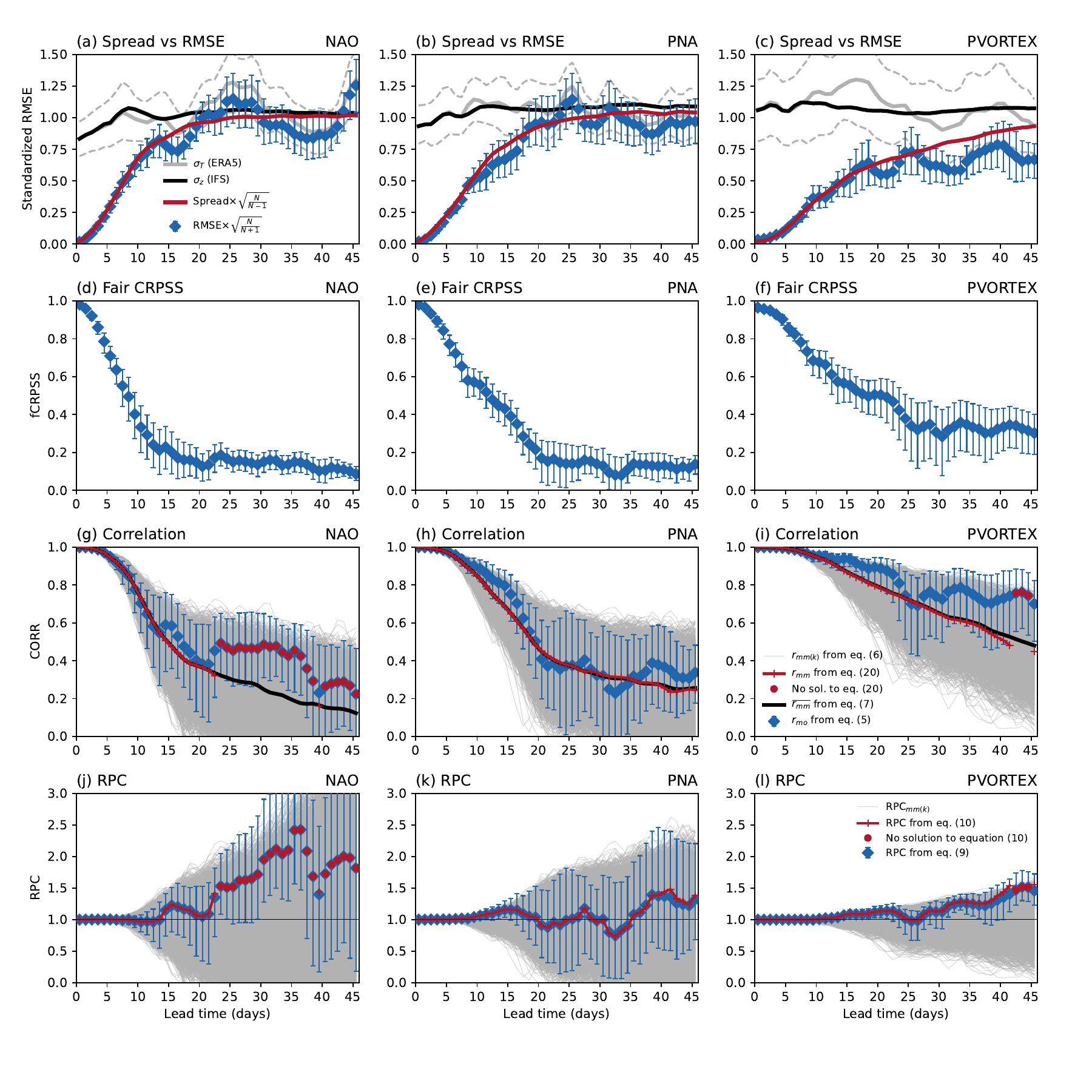}
      \caption{Forecast and verification statistics for three atmosphere circulation indices derived using uncalibrated daily mean anomalies for the period 2001-2020 from Tco319 reforecasts ($M=80$, $N=100$). (a-c) Total anomaly variability in IFS reforecasts ($\sigma_z$) and the ERA5 reanalysis ($\sigma_T$), root-mean-square error of ensemble mean anomaly forecasts (RMSE), and the square root of mean ensemble variance (SPREAD). RMSE and SPREAD are scaled by $\sqrt{\frac{N}{N+1}}$ and $\sqrt{\frac{N}{N-1}}$, respectively, to provide estimates that are unbiased with ensemble size \citep[][]{leutbecher2008ensemble}. (d-f) Fair version of the continuous ranked probability skill score (fCRPSS) calculated as $\textnormal{fCRPSS} = 1 - \frac{\textnormal{fCRPS}}{\textnormal{CRPS}_{\textnormal{clim}}}$, where fCRPS is the fair version of the continuous ranked probability score \citep{ferro2014fair} and CRPS$_{\textnormal{clim}}$ is a reference score derived from the climatological distribution of observed anomalies. (g-i) Correlation between the forecast ensemble mean and observations ($r_{mo}$) and the mean correlation between forecast ensemble mean and an excluded forecast member ($\overline{r_{mm}}$). Grey lines correspond to 10,000 estimates of $r_{mm(k)}$, which represent model-model equivalents of $r_{mo}$ derived by randomly excluding a single member as the `truth' for each start date. The thin red lines correspond to $r_{mm}$ estimated using equation \ref{equation:r_mm_solutions}. Red dots indicate combinations of $r_{mo}$, SER, and VR without solutions to equation \ref{equation:r_mm_solutions}. (j-l) The ratio of predictable components calculated using equations \ref{equation:rpc_corr} and \ref{equation:rpc_solutions}. Red dots correspond to sample combinations of $r_{mo}$, SER, and VR without solutions to equation \ref{equation:rpc_solutions}. Grey lines are the model-model estimates of RPC that correspond to $r_{mm(k)}$ in panels g-i. Error bars or dashed lines represent the 2.5th and 97.5th percentiles of distributions derived by bootstrap resampling (with replacement) from the available start dates 500 times. For clarity, we do not plot error bars for all sample estimates.}

      \label{fig:uncalibrated_verification}
  \end{figure}

To begin, we consider the reliability characteristics of daily mean NAO, PNA, and PVORTEX indices in our large-ensemble Tco319 reforecasts that consist of 80 cases covering the period 2001-2020 (figure \ref{fig:uncalibrated_verification}). In general, there is good agreement between ERA5 and IFS estimates of total NAO variability such that estimates of $\sigma_z$ lie within the 95\% confidence intervals of $\sigma_T$ across all lead times (figure \ref{fig:uncalibrated_verification}a). Similarly, the ensemble spread of NAO forecasts lies within the 95\% confidence intervals of RMSE for almost all lead times. PNA forecasts also show  good agreement between IFS and ERA5 estimates of total variability and a close correspondence between spread and RMSE (figure \ref{fig:uncalibrated_verification}b). Based on these comparisons, daily mean NAO and PNA forecasts seem to satisfy the climatological and ensemble variance reliability criteria described in section \ref{section:reliability} within the tolerance of our estimated sampling uncertainties. In contrast, although PVORTEX forecasts show good agreement between $\sigma_T$ and $\sigma_z$ across all lead times, they seem to become over-dispersive (i.e. spread > RMSE) at lead times greater than 25 days (figure \ref{fig:uncalibrated_verification}c).

For NAO and PNA forecasts, ensemble spread increases smoothly and monotonically with lead time before saturating and converging with estimates of $\sigma_z$. PVORTEX forecasts also show a smooth and monotonic increase in spread with lead time, but it does not saturate within the duration of the 46-day forecasts due to the higher predictability of this stratospheric index. The mean correlation between the forecast ensemble mean and an excluded ensemble member ($\overline{r_{mm}}$) also reduces smoothly with lead time in all three indices due to the gradual loss of predictability at longer time scales (figure \ref{fig:uncalibrated_verification}g-i). In contrast, RMSE, $\sigma_T$, and correlations between forecast ensemble means and observations ($r_{mo}$) exhibit large variations with lead time, which is a consequence of the much larger sampling uncertainty in the verifying observations compared to the 100-member forecast ensemble. The variability in forecast skill with lead time is less evident in the probabilistic continuous ranked probability skill score (CRPSS; figure \ref{fig:uncalibrated_verification}d-f), which measures the skill of the entire forecast distribution relative to a climatological reference forecast.

The evolution of spread, RMSE, CRPSS, and $\overline{r_{mm}}$ with lead time provide a consistent characterization of the relative predictability of the three circulation indices in IFS reforecasts. For example, it takes $\sim$10 days for NAO forecasts to reach a threshold CRPSS value of 0.4. In contrast, PNA and PVORTEX indices are more predictable and reach this threshold value after $\sim$15 and $\sim$25 days, respectively. The order of diagnosed predictability (PVORTEX $>$ PNA $>$ NAO) does not change if timescales are instead diagnosed from threshold values of RMSE, ensemble spread, or $\overline{r_{mm}}$. The exact thresholds and absolute timescales used for this comparison are not critical for diagnosing the relative predictability of each index.

Estimates of predictability derived from $r_{mo}$ are a notable outlier as the NAO is seemingly more predictable than the PNA at some lead times. For PNA forecasts, $\overline{r_{mm}}$ and $r_{mo}$ are generally consistent and thus $\textnormal{RPC} \approx 1$ for all forecast lead times (figure \ref{fig:uncalibrated_verification}k). In contrast, there are notable differences between $\overline{r_{mm}}$ and $r_{mo}$ in NAO and PVORTEX forecasts at lead times greater than 20 days (figures \ref{fig:uncalibrated_verification}j and \ref{fig:uncalibrated_verification}l). In particular, NAO forecasts exhibit an increase in $r_{mo}$ from $\sim$0.40 at day 20 to $\sim$0.46 at day 30 whereas $\overline{r_{mm}}$ decreases from $\sim$0.38 to $\sim$0.27 over the same lead times. These differences between $\overline{r_{mm}}$ and $r_{mo}$ in NAO forecasts result in some lead times when 95\% confidence intervals do not intersect with $\textnormal{RPC}=1$ (e.g. days 31 to 37). Similarly, $r_{mo}$ is significantly higher than $\overline{r_{mm}}$ for some lead times in PVORTEX forecasts (e.g. days 43-46) such that RPC reaches a maximum value of $\sim$1.5. These results for large-ensemble reforecasts during the period 2001-2020 are qualitatively unchanged when start dates are restricted to the winter months (figure \ref{fig:tco319_feb_nov}).

The anomalously high values of RPC for the NAO index at some lead times are seemingly inconsistent with the approximate reliability diagnosed from spread-error and total variance characteristics (figure \ref{fig:uncalibrated_verification}a). This inconsistency is reinforced by the absence of solutions to equation \ref{equation:rpc_solutions} at lead times of 23-39 and 41-46 days, which indicates a qualitatively important role for sampling uncertainties during these periods (figure \ref{fig:uncalibrated_verification}j). In other words, the sample estimates of $r_{mo}$, RPC, SER, and VR for these lead times are statistically paradoxical in the sense that they cannot be interpreted together as valid estimates of the underlying population quantities that would emerge without sampling uncertainty. We also find several lead times in PVORTEX forecasts (e.g. days 43-45) where sample estimates of $r_{mo}$, SER, and VR do not have solutions to equation \ref{equation:rpc_solutions}. In contrast, the RPC and $r_{mm}$ solutions from equations \ref{equation:rpc_solutions} and \ref{equation:r_mm_solutions}, respectively, closely track sample estimates at all lead times for the PNA index. 

The substantial RPC uncertainties at subseasonal lead times in our large-ensemble reforecasts are further illustrated by the empirical distributions of model-model estimates of RPC$_{mm(k)}$, which are tightly clustered around $\textnormal{RPC}=1$ for lead times less than 10 days before diverging due to the impact of sampling variance (figures \ref{fig:uncalibrated_verification}j-l). In fact, point estimates of $r_{mo}$ and RPC lie within the empirical distributions of their model-model equivalents for all three indices and across all lead times (figures \ref{fig:uncalibrated_verification}g-l). Figure \ref{fig:correlation_vs_ens_size}a-c further illustrates $r_{mo}$ and model-model equivalents as a function of ensemble size for each circulation index at a lead time of 35 days. Consistent with figures \ref{fig:uncalibrated_verification}g-h, NAO and PNA estimates of $r_{mo}$ lie within the distribution of $r_{mm(k)}$ estimates for all ensemble sizes (figure \ref{fig:correlation_vs_ens_size}a). In contrast, PVORTEX estimates of $r_{mo}$ at day 35 either match or exceed the maximum value of $r_{mm(k)}$ for all ensemble sizes (figure \ref{fig:correlation_vs_ens_size}c). The high values of $r_{mo}$ and RPC in PVORTEX indices are consistent with the over-dispersion observed at lead times beyond 25 days. 

Despite these substantial uncertainties in RPC and the approximate reliability indicated by other metrics, sample estimates of RPC for the NAO index exceed one at all lead times beyond day 15 in our large-ensemble reforecasts (figure \ref{fig:uncalibrated_verification}j), a behaviour that occurs in only $\sim$2\% of RPC$_{mm(k)}$ realisations. From this evidence it seems extremely unlikely that sample estimates from our large-ensemble reforecasts could be drawn from a perfectly reliable forecasting system with RPC exactly equal to one. However, the absence of solutions to equation \ref{equation:rpc_solutions} for all lead times when $r_{mo}$ diverges from $\overline{r_{mm}}$ also provides strong evidence that sample estimates of $r_{mo}$, SER, and VR cannot be interpreted as converged population quantities as they represent an invalid combination in the large-sample limit. In other words, these daily mean sample estimates of reliability and/or correlation metrics are sufficiently influenced by sampling uncertainty to preclude a naive interpretation of the daily mean reliability characteristics in this dataset. The reliability characteristics of monthly mean NAO indices in our large-ensemble reforecasts, for which sampling uncertainties are significantly reduced, are examined separately in section \ref{section:monthly_nao}.

\begin{figure}[!htbp]
      \centering           
      \includegraphics[width=15cm]{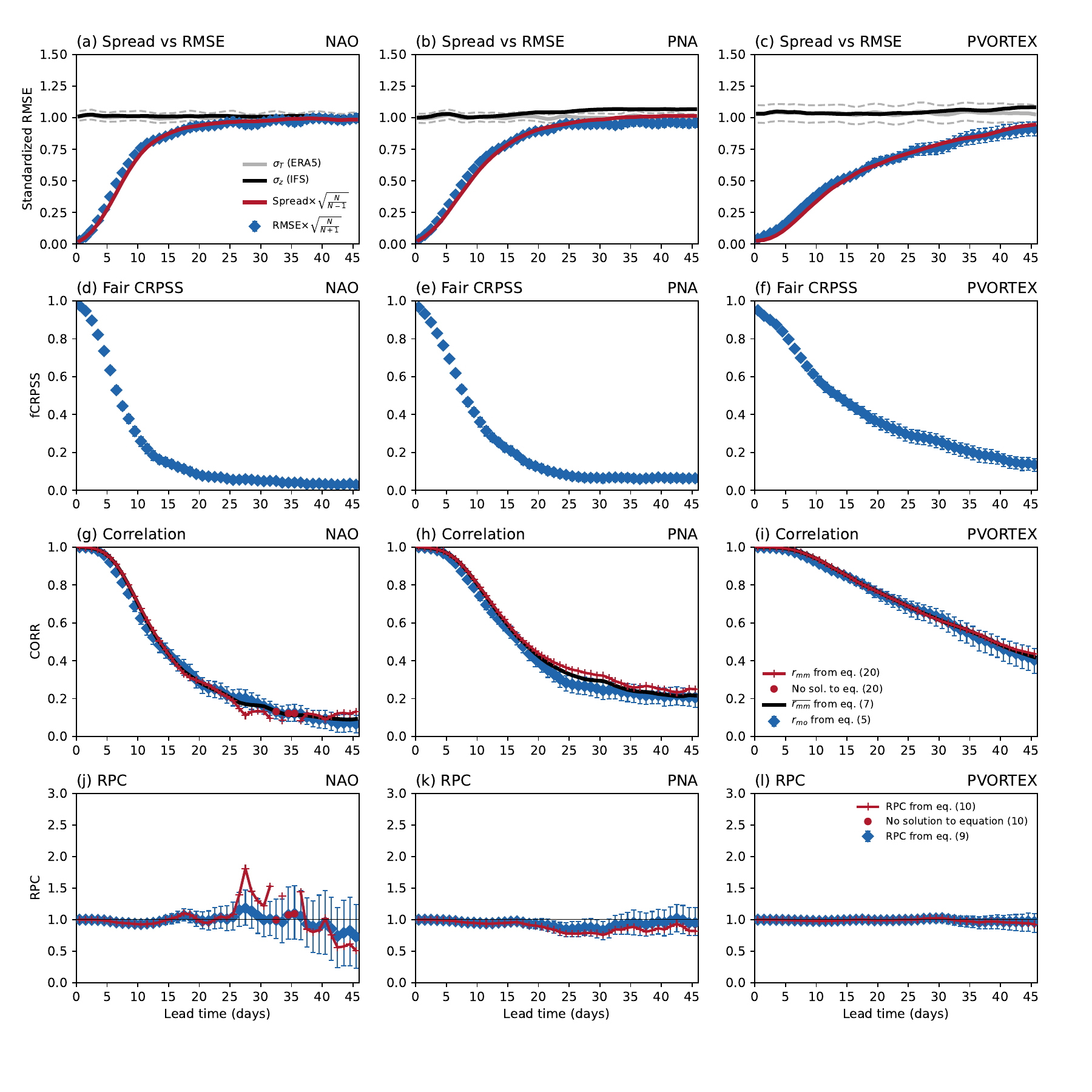}
      \caption{As figure \ref{fig:uncalibrated_verification}, but for the Tco199 reforecasts covering the period 1959-2023 with $M=3120$ and $N=10$.}
      \label{fig:tco199_verification}
  \end{figure}

The reliability characteristics and sampling uncertainties described above for our Tco319 100-member reforecasts are qualitatively and quantitatively extremely similar in our lower-resolution Tco199 100-member reforecasts when evaluated over the same start dates (figure \ref{fig:tco199_100member}). The substantial observational sampling uncertainties in these 100-member reforecasts are greatly reduced in the 10-member Tco199 reforecasts, which are evaluated using 3120 start dates over the period 1959-2023 (figure \ref{fig:tco199_verification}). The reduced observational sampling uncertainties in this extended dataset are evident in the stability of $\sigma_T$ estimates across lead times. Furthermore, all forecast skill estimates reduce smoothly with lead time, and provide a consistent characterisation of the relative predictability of each index such that PVORTEX $>$ PNA $>$ NAO.  Given the evidence from large-ensemble reforecasts, one might expect that systematic model deficiencies (e.g. SNP-like unreliability) would become more evident when evaluated over a much larger sample of start dates. However, we find that NAO and PVORTEX indices evaluated over the period 1959-2023 are remarkably well-calibrated such that $\textnormal{RPC} \approx 1$ across all lead times and all but three NAO lead times have valid solutions to equation \ref{equation:rpc_solutions} (figure \ref{fig:tco199_verification}). This evaluation of Tco199 reforecasts for the period 1959-2023 is unchanged when analysis is restricted to start dates for the extended winter season (figure \ref{fig:tco199_extended_winter}).

\section{Calibrated daily mean forecasts}
\label{section:results_calibrated}

\subsection{Direct calibration of circulation indices}
\label{section:results_direct_calibration}
This section evaluates the reliability and signal-to-noise characteristics of daily mean NAO, PNA, and PVORTEX indices from our Tco319 large-ensemble reforecasts after application of an unbiased member-by-member calibration, which simultaneously enforces the climatological reliability and ensemble variance reliability criteria described in section \ref{section:statistics}. The estimated calibration parameters $\alpha$ and $\beta$ modify the ensemble mean (i.e. the predictable \textit{signal}) and perturbations from the ensemble mean (i.e. the unpredictable \textit{noise}), respectively. Parameters are estimated separately for each lead time and start month. We do not make any separation between training and verification data when estimating calibration parameters as the intention is to understand the statistical properties of this set of reforecasts rather than optimise the skill of a real-time forecast system.

\begin{figure}[!htbp]
      \centering           
      \includegraphics[width=15cm]{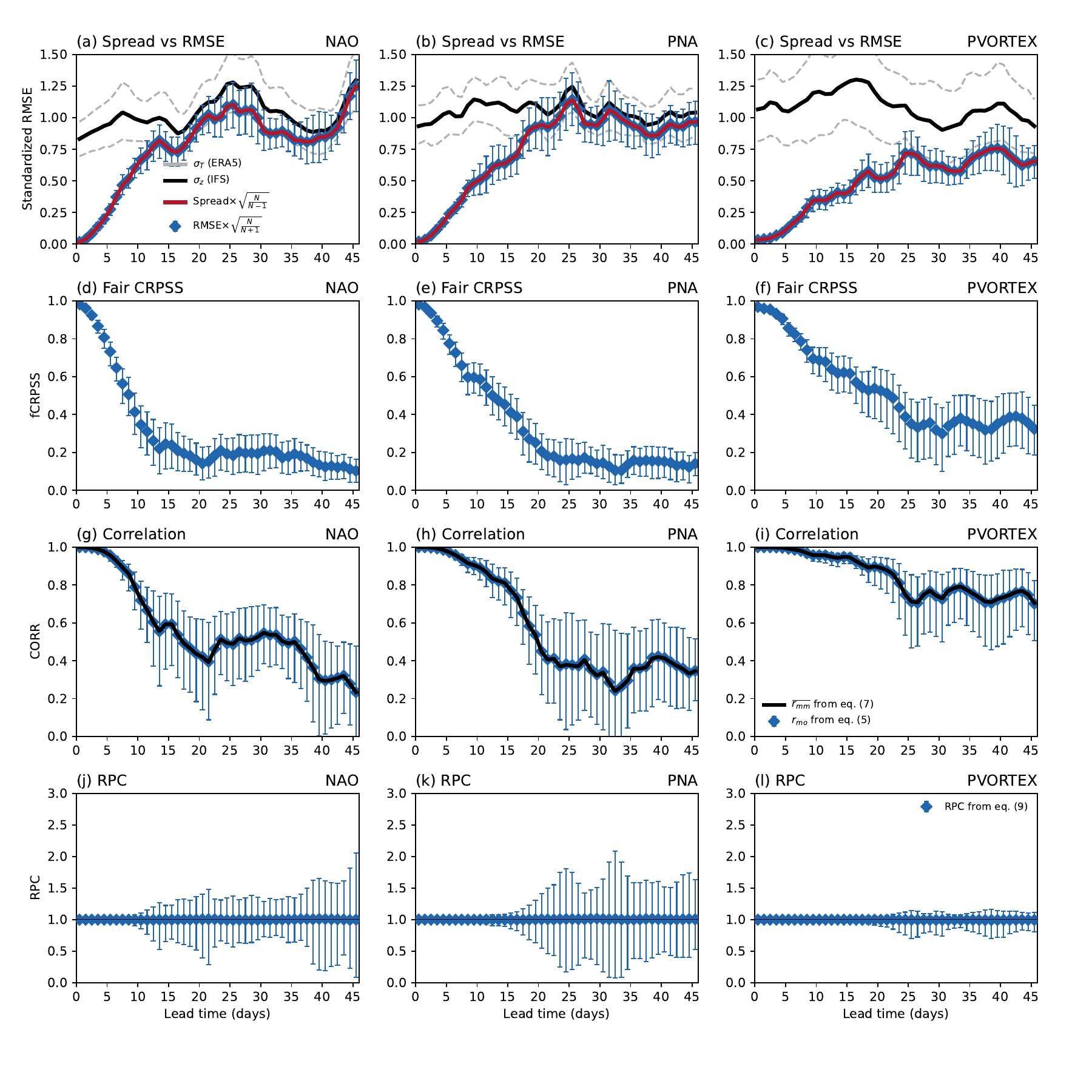}
      \caption{As figure \ref{fig:uncalibrated_verification}, but for calibrated circulation indices.}
      \label{fig:calibrated_index_verification}
  \end{figure}

The results of calibrating each forecast index are summarised in figure \ref{fig:calibrated_index_verification}. As expected, the in-sample reliability calibration enforces the constraints that SER = 1 and VR = 1 (figure \ref{fig:calibrated_index_verification}a-c). Calibration also modifies $\overline{r_{mm}}$ to match $r_{mo}$ such that $\textnormal{RPC}=1$ at all lead times in all three circulation indices (figure \ref{fig:calibrated_index_verification}g-i). In spite of the `perfect' RPC values and substantial changes to $\overline{r_{mm}}$, $\sigma_z$, and ensemble spread, calibration has a limited impact on forecast skill diagnosed using RMSE, $r_{mo}$, and CRPSS (figure \ref{fig:calibrated_index_verification}). Furthermore, the ensemble spread of calibrated forecasts no longer increases smoothly and monotonically with lead time as it is forced to inherit the variations with lead time that are present in RMSE. Similarly, estimates of $\sigma_z$ and $\overline{r_{mm}}$ derived from calibrated forecasts also inherit the large variations with lead time that are present in $\sigma_T$ and $r_{mo}$, respectively. Enforcing the constraint that $\textnormal{RPC}=1$, $\textnormal{VR}=1$, and $\textnormal{SER}=1$ at each lead time thus leads to overfitting to the available observations, such that sample statistics from calibrated forecasts inherit the large sampling uncertainties present in the observations.

\begin{figure}[!htbp]
      \centering           
      \includegraphics[width=15cm]{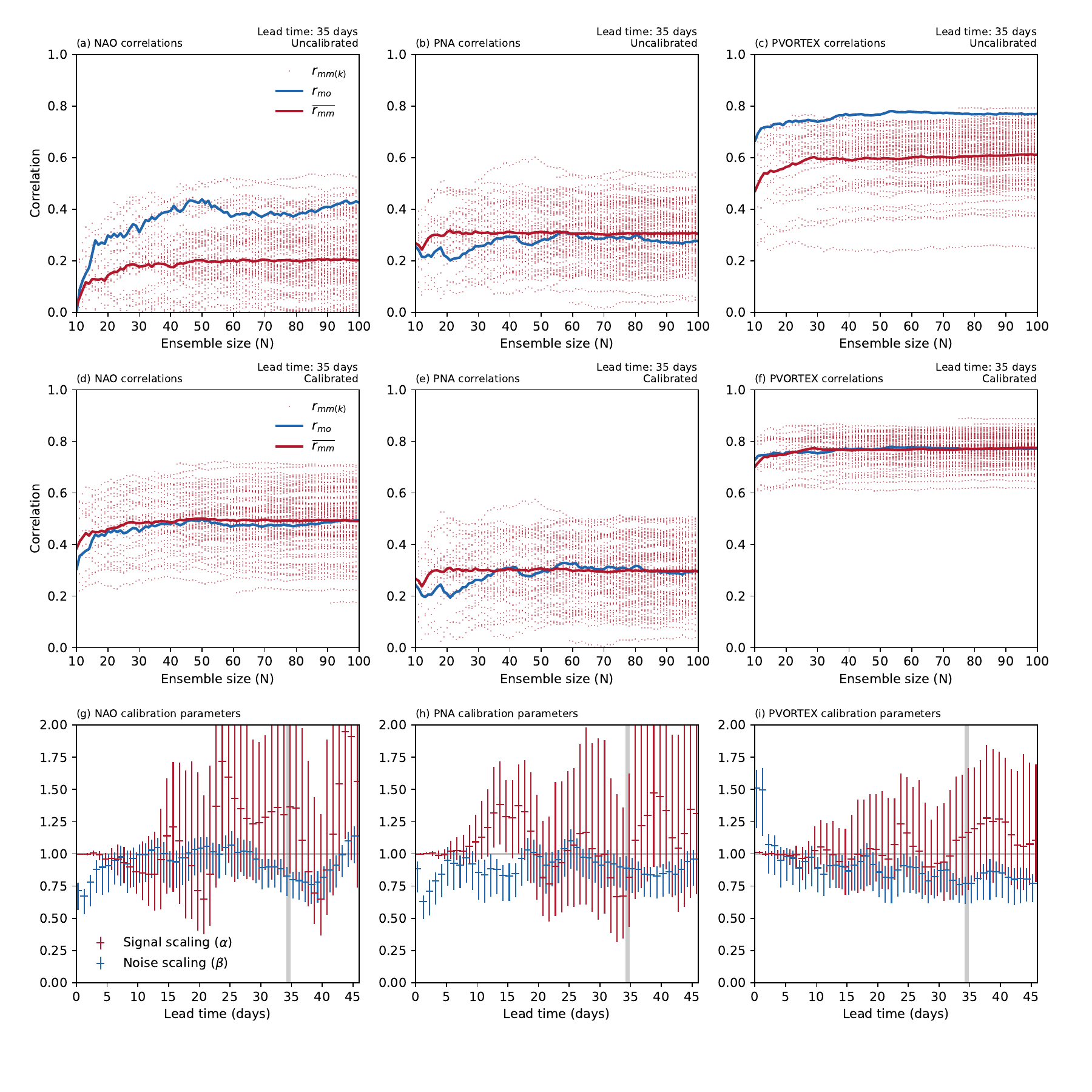}
      \caption{(a-c) Correlations vs ensemble size for circulation indices calculated using uncalibrated daily mean anomalies for the period 2001-2020 from Tco319 reforecasts ($M=80$, $N=100$), where $r_{mo}$ is the correlation between the forecast ensemble mean and ERA5, $r_{mm(k)}$ is the correlation between a forecast ensemble mean and an excluded ensemble member, and $\overline{r_{mm}}$ is the mean of $N+1$ estimates of $r_{mm(k)}$. (d-f) As above, but for indices calibrated using an unbiased member-by-member approach that simultaneously enforces climatological reliability and ensemble variance reliability (see section \ref{section:calibration}). (g-i) Mean of calibration parameters $\alpha$ and $\beta$ (see equations \ref{equation:alpha} and \ref{equation:beta}) used in panels d-f. Uncertainties in parameter values are estimated using a bootstrap resampling approach whereby average calibration parameters are calculated 500 times using randomly selected (with replacement) start years. Error bars represent the 2.5th and 97.5th percentiles of the resulting distributions. The vertical grey bar indicates the lead time plotted in panels a-f.}
      \label{fig:correlation_vs_ens_size}
  \end{figure}

Figure \ref{fig:correlation_vs_ens_size}d-f shows estimates of $r_{mo}$, $\overline{r_{mm}}$, and $r_{mm(k)}$ vs ensemble size from calibrated index forecasts for a lead time of 35 days. In a perfectly reliable ensemble, $r_{mm(k)}$ and $r_{mo}$ can be considered drawn from the same underlying probability distribution and their values will converge with $\overline{r_{mm}}$ when sample statistics are evaluated over many independent start dates (see discussion in section \ref{section:uncertainty}). However, despite the perfect agreement between $\overline{r_{mm}}$ and $r_{mo}$ across all lead times (for $N=99$), the calibrated forecasts still exhibit a large spread in estimates of $r_{mm(k)}$ (figure \ref{fig:correlation_vs_ens_size}d-f). This is inconsistent with our expectations of a perfectly reliable ensemble and is evidence that the `perfect' RPC values in our finite set of forecasts can only be achieved through some degree of overfitting.

Despite the overfitting issues discussed above, it is still instructive to evaluate the calibration parameters $\alpha$ and $\beta$ and their associated uncertainties as a function of lead time (figure \ref{fig:correlation_vs_ens_size}g-i). Crucially, we do not find statistically robust evidence for a consistent underestimation of the magnitude of predictable signals (i.e. $\alpha > 1$) for any of the three circulation indices. For example, estimates of $\alpha$ for NAO forecasts vary substantially with lead time between values of $\sim$0.6 and $\sim$1.9 with large uncertainty estimates that overlap $\alpha=1$. In contrast, estimates of $\beta$ have much smaller sampling uncertainties with several features that are worthy of comment. Firstly, short-range NAO and PNA forecasts have $\beta<1$, which is indicative of over-dispersion at these lead times. In contrast, short-range PVORTEX forecasts have $\beta>1$, which is indicative of under-dispersion. However, the absolute values of spread are very small at these lead times and thus differences between spread and error are not evident in figure \ref{fig:uncalibrated_verification}a-b. PNA and NAO forecasts also exhibit other periods with $\beta<1$, but these generally correspond to lead times when RMSE and $\sigma_T$ are reduced compared to surrounding lead times, which is indicative of observational sampling uncertainty. Lastly, PVORTEX forecasts exhibit a seemingly statistically robust $\beta<1$ at lead times greater than 20 days (figure \ref{fig:correlation_vs_ens_size}i). This is consistent with the over-dispersion (i.e. spread $>$ RMSE) at lead times greater than 25 days that is associated with RPC $>1$ (figure \ref{fig:uncalibrated_verification}).

\subsection{Indirect calibration of circulation indices}
\label{section:results_anomaly_calibration}
We also evaluate the impact of an indirect calibration approach applied to our Tco319 large-ensemble reforecasts, whereby forecast anomalies are calibrated separately for each grid-point, start month, and lead time prior to calculating forecast indices. This allows us to evaluate both the reliability and signal-to-noise characteristics of the circulation indices together with other aspects of the circulation, such as tropical-extratropical teleconnections. 

The impact of indirect anomaly calibration (figure \ref{fig:calibrated_gridpoint}) is similar, but not identical, to the impact of direct calibration of circulation indices (figure \ref{fig:calibrated_index_verification}). There is improved agreement between both (i) spread and RMSE and (ii) $\sigma_z$ and $\sigma_T$, which comes at the cost of large variations with lead time as discussed in section \ref{section:results_direct_calibration}. In addition, there is closer agreement between $\overline{r_{mm}}$ and $r_{mo}$ such that $\textnormal{RPC} \approx 1$ within our estimated sampling uncertainties at all lead times in all three circulation indices (figure \ref{fig:calibrated_gridpoint}). The differences between calibration methods are a consequence of the covariance between grid points, which are not accounted for when calibrating grid-points independently. For example, it is possible for grid points to individually have perfect variances, but the variance of their sum can be incorrect if there are errors in the correlation between grid-points.

In spite of this `imperfect' indirect calibration and the overfitting issues discussed in section \ref{section:results_direct_calibration}, these calibrated anomalies provide an opportunity to evaluate other properties of the atmospheric circulation when $\textnormal{RPC} \approx 1$. \citet{roberts2023euro} recently demonstrated that ECMWF reforecasts with IFS cycle 47R3 accurately simulate wintertime Euro-Atlantic regime structures, frequencies, and transition probabilities, at subseasonal lead times. However, they emphasised that IFS reforecasts underestimate the response of the NAO to the Madden-Julian oscillation (MJO) and fail to reproduce the modulation of MJO-NAO teleconnections by El Ni\~{n}o-Southern Oscillation (ENSO). These conditional errors were attributed to deficiencies in the representation of tropical-extratropical teleconnections, which have been identified in previous IFS cycles and other subseasonal forecast systems \citep[e.g. ][]{vitart2017madden}. Importantly, underestimation of tropical-extratropical teleconnection signals such that forecasts do not fully exploit the response of the extratropics to predictable intraseasonal variability in the tropics is one of the proposed physical interpretations for the SNP in seasonal forecasts \citep{garfinkel2022winter, scaife2018signal}. However, as emphasised in section \ref{section:calibration}, RPC is an unconditional reliability metric and calibration that enforces $\textnormal{VR} = \textnormal{SER} = \textnormal{RPC} = 1$ when averaged over all start dates does not necessarily require or imply conditional reliability. It is thus possible to satisfy these constraints in the extratropics without requiring perfect statistical consistency within subsets corresponding to different states of a tropical forcing (e.g. different MJO or ENSO phases). 

\begin{figure}[!htbp]
      \centering           
      \includegraphics[width=15cm]{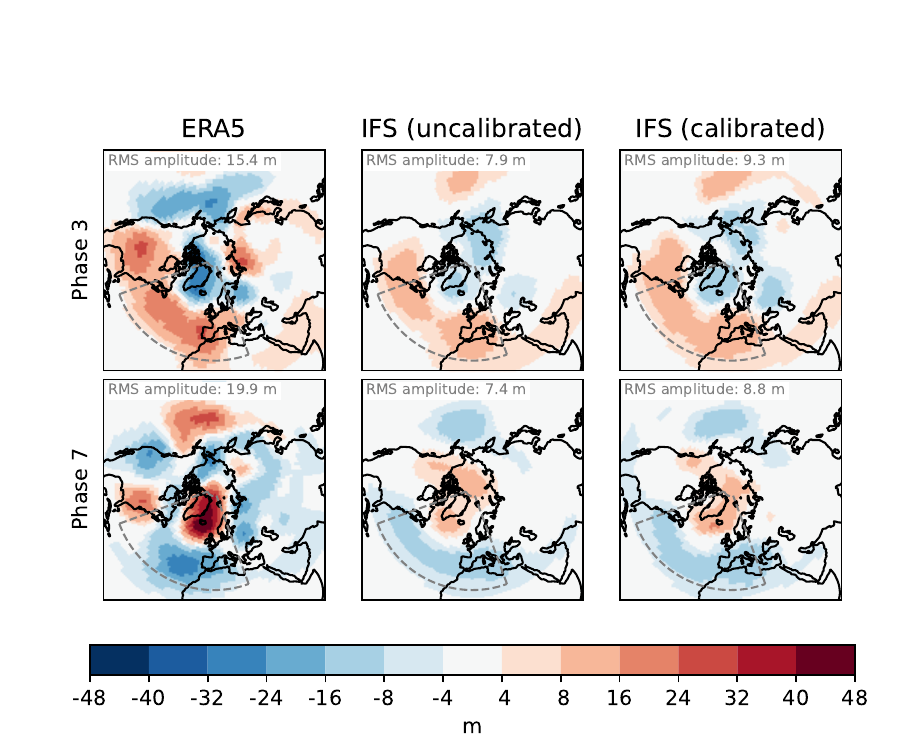}
      \caption{Composite means of 500 hPa geopotential height anomalies 15 days after the specified MJO phase for the period 2001-2020 based on daily mean ERA5 and Tco319 IFS reforecasts with $M=80$ and $N=100$. Calibrated composites are constructed using MJO indices derived from forecast anomalies that have been calibrated separately for each grid-point, start month, and lead time as described in section \ref{section:results_anomaly_calibration}. Contributing data are selected using the MJO phase calculated separately in each forecast member and weak amplitude events (i.e. $\sqrt{\textnormal{RMM1}^2 + \textnormal{RMM2}^2} < 1$) are excluded from the composite calculation. All forecast lead times are considered together (i.e. composite means are constructed from forecast anomalies corresponding to days 16-46 using MJO phases identified during days 1-31). ERA5 data are subsampled to exactly match the available forecast data. Annotated values indicate the area-weighted root-mean-square amplitude of the composite patterns in the indicated Euro-Atlantic domain.}
      \label{fig:mjo_z500_composites}
  \end{figure}

\begin{figure}[!htbp]
      \centering           
      \includegraphics[width=15cm]{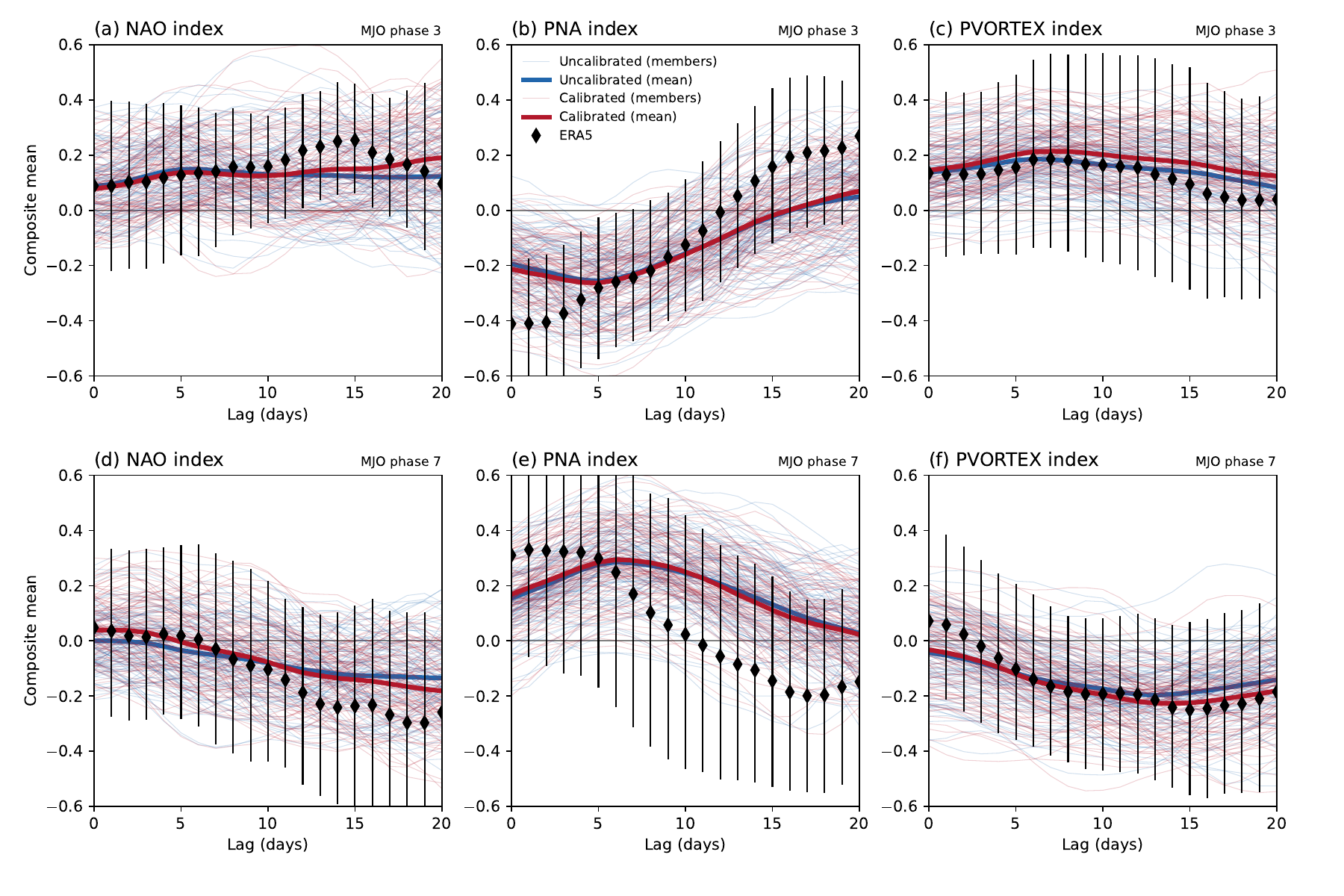}
      \caption{Lagged composites of daily circulation indices (NAO, PNA, PVORTEX) conditioned on the phase of the MJO for the period 2001-2020 based on daily mean ERA5 and Tco319 IFS reforecasts with $M=80$ and $N=100$. As in figure \ref{fig:mjo_z500_composites}, contributing data are selected using the MJO phase calculated separately in each forecast member and weak amplitude events (i.e. $\sqrt{\textnormal{RMM1}^2 + \textnormal{RMM2}^2} < 1$) are excluded from the composite calculation. Uncertainties in ERA5 composites are estimated by bootstrap resampling (with replacement) from the available start dates such that error bars represent the 2.5th and 97.5th percentiles of the resulting distribution. Blue and red lines represent composites constructed from uncalibrated and calibrated forecast data, respectively. Calibrated composites are constructed using indices derived from forecast anomalies that have been calibrated separately for each grid-point, start month, and lead time as described in section \ref{section:results_anomaly_calibration}. Bold red/blue lines represent composites constructed using 100 forecast members. Thin red/blue lines represent composites constructed using a single member from each forecast start date.}
      \label{fig:mjo_index_composites}
  \end{figure}

Our evaluation of ERA5 teleconnections (figures \ref{fig:mjo_z500_composites} and \ref{fig:mjo_index_composites}) is qualitatively consistent with previous studies that have described the impact of the MJO on the NAO, PNA, and PVORTEX \citep[e.g.][]{cassou2008intraseasonal,lin2009observed,  garfinkel2012observed,seo2012global, garfinkel2014impact, barnes2019tropospheric, lee2019enso, wang2020mjo, roberts2023euro}. In particular, ERA5 geopotential height anomalies in the Euro-Atlantic sector that occur 15 days after MJO phases 3 and 7 (figure \ref{fig:mjo_z500_composites}) project onto the positive and negative phases of the NAO, respectively (figure \ref{fig:mjo_index_composites}). Uncalibrated IFS reforecasts also simulate an NAO response to the MJO, but the lagged composites constructed from 100 forecast members are much weaker than estimates based on ERA5 data (figures \ref{fig:mjo_z500_composites} and \ref{fig:mjo_index_composites}). In the Euro-Atlantic domain, the RMS amplitude of gepotential height anomalies 15 days after MJO phases 3 and 7 are higher in calibrated forecasts (9.3 m and 8.8 m) compared to uncalibrated forecasts (7.9 and 7.4 m), but both are substantially lower than ERA5 estimates (15.4 m and 19.9 m). However, consistent with our discussion of ERA5-based sample statistics, there is considerable sampling uncertainty in NAO, PNA, and PVORTEX composites constructed from daily data such that 100-member IFS composites are within the 95\% confidence limits of ERA5 composites for all indices and MJO phases/lags (figure \ref{fig:mjo_index_composites}). Similarly, ERA5-based composites lie within the distribution of uncalibrated IFS estimates based on a single member from each forecast start date (figure \ref{fig:mjo_index_composites}). From this comparison it is clear that more start dates and/or longer composite averaging periods are required to robustly detect differences between IFS and ERA5 MJO teleconnections. 

Nevertheless, the important result for this study is that MJO-index teleconnections are very similar in calibrated and uncalibrated forecasts (figure \ref{fig:mjo_index_composites}). For example, the magnitude of the NAO index in the 15-20 days following MJO phase 3/7 is slightly higher in calibrated forecasts, but this difference is small compared to the uncertainty in the ERA5-based composites. In general, the detailed representation of MJO-index teleconnections in these reforecasts seems to be independent of the presence or absence of SNP-like behaviour in the underlying index. For example, the largest discrepancy between ERA5 and forecast MJO composites is for the PNA, for which $\overline{r_{mm}}$ and $r_{mo}$ are generally consistent and thus $\textnormal{RPC} \approx 1$ for all forecast lead times. We expect improvements in the representation of tropical-extratropical teleconnections to be associated with improvements in extratropical skill. However, perfect teleconnections and conditional reliability are not required for $\textnormal{RPC} \approx 1$, nor does calibration necessarily lead to their improvement. The only condition for RPC $\rightarrow$ 1 as $M,N \rightarrow \infty$ is that forecasts exhibit unconditional reliability and satisfy the climatological and ensemble variance criteria described by equations \ref{equation:variances} and \ref{equation:spread_error}.

\section{Monthly mean NAO indices}
\label{section:monthly_nao}

The previous sections highlighted the challenges of interpreting reliability and signal-to-noise characteristics of daily-mean circulation indices in our large-ensemble reforecasts due to the large sampling uncertainties inherent in daily observational data. Here, we examine how temporal averaging influences these characteristics by analysing monthly-mean NAO forecasts, obtained by averaging over lead times of 1-30 and 16-45 days. In addition to the EOF-based NAO index derived from Z500 anomalies considered in sections \ref{section:results_uncalibrated} and \ref{section:results_calibrated}, we evaluate two other NAO indices derived from mean sea-level pressure (MSLP): an EOF-based MSLP index and a station-based index defined as the MSLP difference between Lisbon and Reykjavík (see section \ref{section:indices} for details). To assess seasonality, we also stratify our analysis by winter and summer start dates.

\begin{figure}[!htbp]
      \centering           
      \includegraphics[height=18cm]{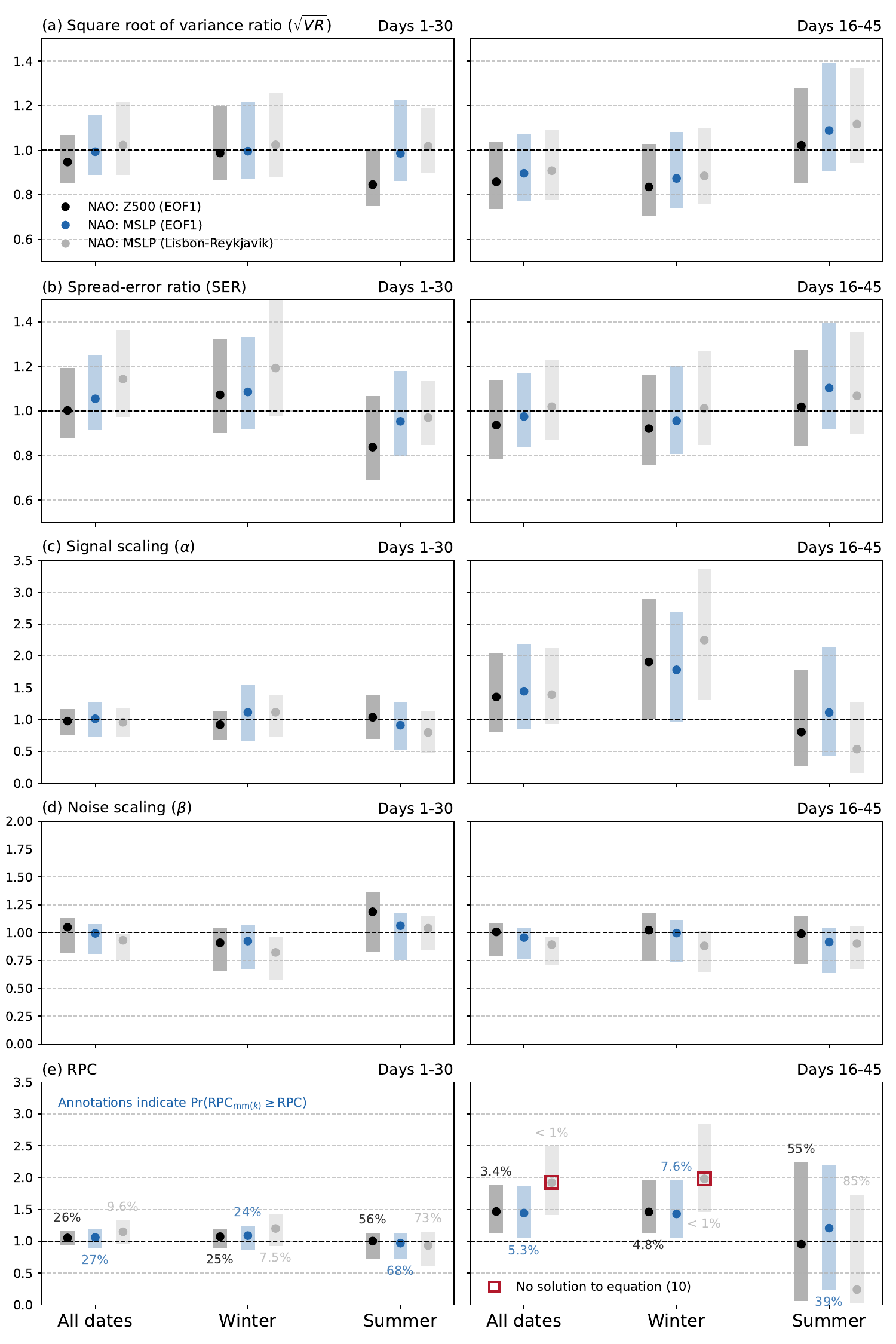}
      \caption{Monthly mean NAO verification statistics for Tco319 reforecasts covering the period 2001-2020 with $N=100$ and $M=80$ for all start dates or $M=40$ when stratified by season. (a) Square root of the total variance ratio ($\sqrt{\textnormal{VR}}$) for time-averaged NAO indices corresponding to lead times of 1-30 days and 16-45 days. Different colours correspond to different definitions of the NAO index, as indicated in the legend. Time-averaged NAO statistics are for all start dates together and also separately for summer (May 1st and August 1st) and winter (November 1st and February 1st) periods. Confidence intervals represent the 2.5th and 97.5th percentiles of bootstrap distributions derived by resampling (with replacement) from the available start dates 500 times. (b) As above but for the spread-error ratio (SER; equation \ref{equation:spread_error}). (c, d) As above but for unbiased reliability calibration parameters calculated following section \ref{section:calibration}. (e) As above, but for the RPC calculated following equation \ref{equation:rpc_corr}. Text annotations indicate the percentage of model-model estimates of RPC$_{mm(k)}$ that exceed the sample estimate of RPC. Red boxes indicate RPC estimates without solutions to equation \ref{equation:rpc_solutions}.}
      \label{fig:monthly_nao_verification}
  \end{figure}

We first consider the reliability characteristics of monthly mean NAO indices in our Tco319 large-ensemble reforecasts covering the period 2001-2020. All three NAO indices averaged over days 1-30 satisfy reliability criteria in both summer and winter, as indicated by 95 \% confidence intervals that intersect with SER = 1, VR = 1, and RPC = 1 (figure \ref{fig:monthly_nao_verification}). Furthermore, NAO RPC estimates lie well within the empirical distributions of model-model equivalents (figure \ref{fig:monthly_nao_verification}e), and the diagnosed reliability calibration parameters are mostly indistinguishable from 1. The exception is the station-based MSLP index, which shows $\beta < 1$, indicative of over-dispersion for this index when averaged over the first month. This behaviour is consistent with its slightly elevated RPC and SER values relative to the EOF-based indices.

For the longer averaging period (16-45 days), all three indices derived from large-ensemble reforecasts satisfy climatological and ensemble-variance reliability, with 95 \% confidence intervals intersecting SER = 1 and VR = 1. However, all-year and wintertime RPC estimates exceed one for all three indices, and 95\% confidence intervals do not intersect RPC = 1 (figure \ref{fig:monthly_nao_verification}). For the EOF-based NAO indices, this inconsistency between reliability metrics is not statistically paradoxical as defined in section \ref{section:uncertainty} and likely reflects correlated uncertainties between SER and VR such that bootstrap samples are clustered in regions of SER-VR space corresponding to RPC > 1 (see figure \ref{fig:rpc_vs_ser_vr}). In contrast, summer start dates are generally reliable and confidence intervals for all three indices include RPC = 1.

The EOF-based NAO indices derived from Z500 and MSLP anomalies exhibit nearly identical statistical behaviour. Their all-year and winter RPC values are centred near 1.5, with 95 \% confidence intervals covering the range 1.1--1.9. The probability that model-model RPC$_{mm(k)}$ exceeds the corresponding observation-based sample estimate ranges from 3.4 \% to 7.6 \%, depending on index and season. These elevated RPC values correspond to ensemble mean calibration factors of $\alpha \approx 1.4$ (all start dates) and $\alpha \approx 1.8$ (winter only). However, the lower bounds of our 95 \% confidence intervals for EOF-based NAO indices either intersect or nearly intersect $\alpha = 1$, indicating marginal significance. The highest values correspond to the wintertime Z500 EOF-based index, for which the 95 \% interval covers the range 1.01--2.91.

The station-based MSLP NAO index yields systematically higher RPC and $\alpha$ values than either EOF-based index, with the probability of RPC$_{mm(k)} \geq \textnormal{RPC}$ below 1\% in both all-year and winter subsets. However, for this index, the sample combination of reliability and correlation metrics is statistically paradoxical such that there are no valid solutions to equation \ref{equation:rpc_solutions}. In other words, the observed combinations of $r_{mo}$, RPC, SER, and VR would imply negative or complex-valued variances in the large-sample limit, indicating that at least one of these sample estimates is substantially contaminated by sampling noise and precluding their collective interpretation as converged population quantities.

\begin{figure}[!htbp]
      \centering           
      \includegraphics[height=12cm]{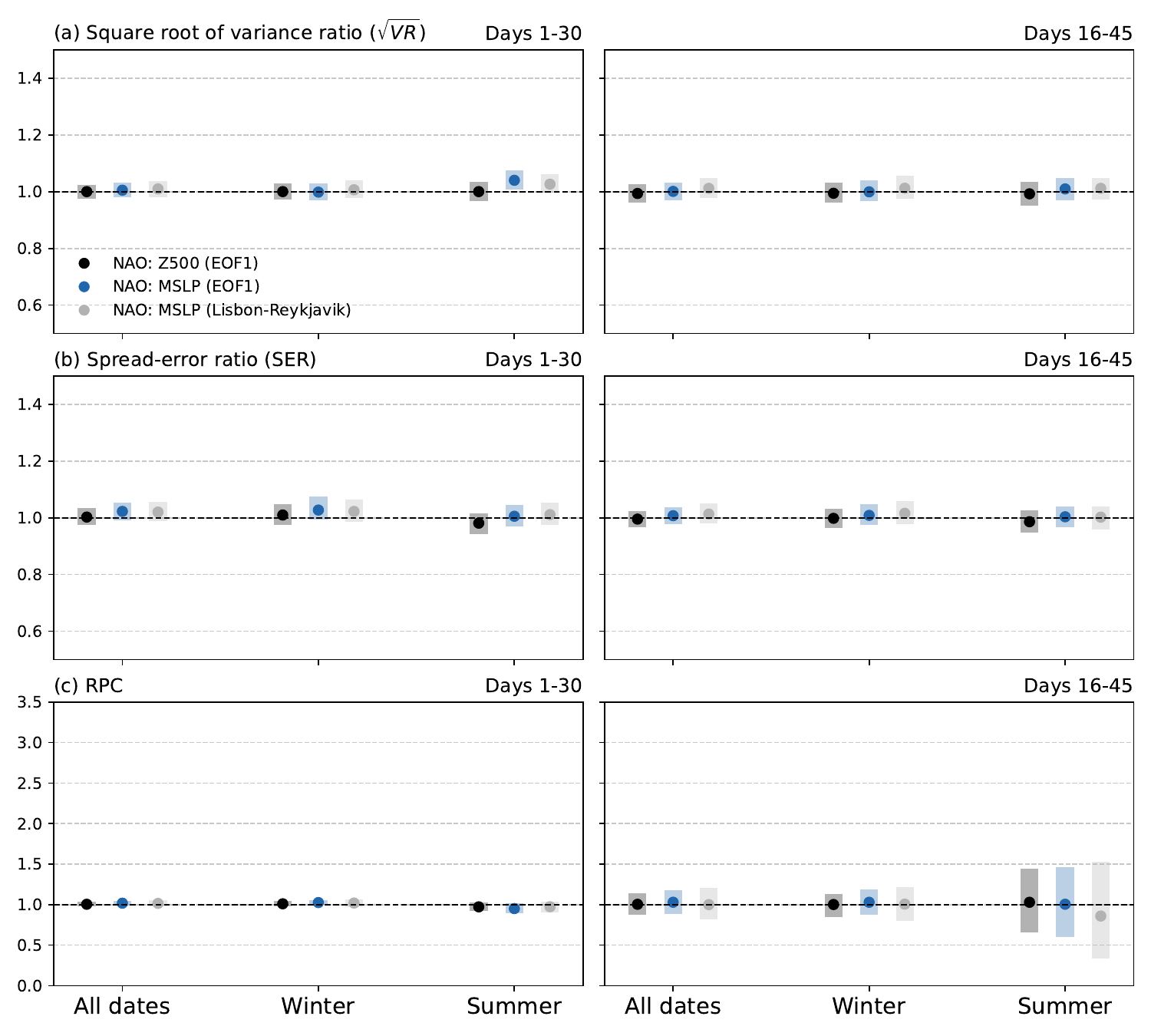}
      \caption{As figure \ref{fig:monthly_nao_verification} without the calibration parameters, but for Tco199 reforecasts covering the period 1959-2023 with $N=10$ and $M=3120$ for all start dates or $M=1560$ when separated into extended summer (April-September) or extended winter (October-March) seasons.}
      \label{fig:monthly_nao_verification_1959_2023}
  \end{figure}

The reliability characteristics and sampling uncertainties of our 100-member Tco319 and Tco199 reforecasts are qualitatively and quantitatively extremely similar when evaluated for the same start dates over the period 2001-2020 (figure \ref{fig:tco199_100member}). In contrast, our extended dataset of Tco199 reforecasts spanning 1959-2023 has substantially reduced observational sampling uncertainties and more consistency in reliability characteristics evaluated for different seasons and lead times. These reforecasts are well-calibrated such that SER=1, VR=1, and RPC=1 within our estimated 95\% confidence intervals (figure \ref{fig:monthly_nao_verification_1959_2023}) for all three NAO definitions and seasons, with one exception. Furthermore, all combinations of SER, VR, and $r_{mo}$ have solutions to equation \ref{equation:rpc_solutions}. The only minor deviation from reliability occurs for the EOF-based MSLP index during days 1-30 during the summer, where the lower bound of the VR confidence interval lies just above one. Otherwise, subseasonal IFS reforecasts evaluated over the extended period 1959-2023 satisfy our unconditional reliablilty criteria for monthly mean NAO variability during both extended summer and extended winter seasons.

\section{Discussion and conclusions}
\label{section:conclusions}

In this study we have illustrated the relationship between SNP-like behaviour and other metrics of ensemble reliability using a general closed-form algebraic expression for RPC in terms of $r_{mo}$, SER, and VR (figure \ref{fig:rpc_vs_ser_vr}). In particular, we emphasized that unbiased estimates of RPC, SER, and VR measure different aspects of statistical consistency with observations, but must be mutually consistent. Physical constraints on the admissible solutions to equation \ref{equation:rpc_solutions} (i.e. real-valued and non-negative variances) provide a mechanism to identify statistically paradoxical sample estimates of reliability and correlation metrics that correspond to combinations that are not possible without sampling uncertainty. We propose using these constraints to discriminate between sample estimates of RPC that are statistically paradoxical due to sampling uncertainties and those that provide more robust evidence for unreliability associated with ensemble mean signals that are too weak. Long-range forecasting systems that predict anomalies in seasonal-to-decadal means may be more vulnerable to such sampling effects due to the limited availability of independent start dates for verification (e.g. figure \ref{fig:rpc_vs_ser_vr}e-f).

In section \ref{section:results_uncalibrated}, we evaluated the forecast skill, reliability characteristics, and signal-to-noise properties of three large-scale atmospheric circulation indices in 100-member subseasonal reforecasts for 80 start dates covering the period 2001-2020. Daily mean NAO, PNA, and PVORTEX forecasts generally satisfied climatological reliability (equation \ref{equation:variances}) and ensemble variance reliability (equation \ref{equation:spread_error}) criteria within the tolerance of our estimated sampling uncertainties. Nevertheless, daily mean NAO forecasts exhibited anomalously high RPC values for some subseasonal lead times (figure \ref{fig:uncalibrated_verification}j). These lead times also coincided with paradoxical combinations of correlation and reliability metrics without solutions to equation \ref{equation:rpc_solutions}, indicating an important role for sampling uncertainties. Nevertheless, when the same forecasts were averaged to produce monthly mean wintertime NAO indices (section \ref{section:monthly_nao}), they exhibited more robust evidence for unreliability characterised by weak predictable signals (RPC $\approx$ 1.5, $\alpha \approx$ 1.8 with 95\% confidence intervals of 1.1--1.9 and 1.0--2.9 respectively), suggesting that the SNP-like behaviour observed in daily data during the same period is not solely attributable to sampling artefacts. However, we also contrasted this result from large-ensemble reforecasts for the period 2001-2020 with evidence from 10-member IFS reforecasts covering the period 1959-2023. In this extended dataset, observational sampling uncertainties were substantially reduced (3120 vs. 80 start dates) and daily NAO indices were found to be well-calibrated with RPC $\approx$ 1 across all subseasonal lead times (figure \ref{fig:tco199_verification}).

In section \ref{section:results_calibrated}, we demonstrated that SNP-like behaviour in daily mean data from our large-ensemble reforecasts could be eliminated by application of an unbiased member-by-member calibration, which produces ensemble forecasts that exactly satisfy the climatological reliability and unbiased ensemble variance reliability conditions described in section \ref{section:reliability}. However, for the NAO, this appeared to be achieved through overfitting such that sample statistics from calibrated forecasts inherited the large sampling uncertainties present in the observations and thus showed large variations with lead time. Consistent with this interpretation, we did not find strong evidence for a systematic underestimation of the magnitude of predictable signals (i.e. $\alpha > 1$) for any of the three circulation indices. For example, estimates of $\alpha$ for NAO forecasts varied substantially with lead time between values of $\sim$0.6 and $\sim$1.9 with large uncertainty estimates overlapping $\alpha=1$. In addition, tropical-extratropical MJO teleconnections were found to be very similar in calibrated and uncalibrated forecasts (figures \ref{fig:mjo_z500_composites} and \ref{fig:mjo_index_composites}). The representation of subseasonal MJO teleconnections in these reforecasts thus appeared independent of the presence or absence of SNP-like behaviour in the underlying dataset. Based on this evaluation, we infer that improvements in the representation of tropical-extratropical teleconnections may be important for future advances in subseasonal forecast skill, but such improvements are not a prerequisite for unconditional reliability averaged over all start dates or eliminating SNP-like behaviour in extratropical circulation indices at subseasonal lead times.

In section \ref{section:monthly_nao}, we examined the impact of seasonality and temporal averaging on the reliability and signal-to-noise characteristics of three different monthly mean NAO indices. In our large-ensemble reforecasts covering the period 2001-2020 we found that averaging over days 1-30 yielded seemingly reliable forecasts for both summer and winter seasons. When averaged over days 16-45, these same forecasts appeared reliable with respect to SER and VR but exhibited elevated RPC values during winter. For the EOF-based indices, these results suggested underestimation of the predictable component by roughly 45\% (i.e. $\alpha=1.8$) in 30-day-mean NAO variability for winter start dates during the period 2001-2020. The station-based NAO index, by contrast, appeared more susceptible to sampling effects as indicated by the absence of valid solutions to equation \ref{equation:rpc_solutions}. For this index, sample estimates of reliability and correlation metrics could not be interpreted collectively as population quantities that would be achieved in the limit $M,N\to \infty$. As for daily indices, we contrasted this result with evidence from extended reforecasts covering the period 1959-2023, which showed that monthly mean NAO forecasts are seemingly reliable (SER=1, VR=1, RPC=1 within 95 \% confidence intervals) for all three NAO definitions and both extended summer and winter seasons (figure \ref{fig:monthly_nao_verification_1959_2023}), with all combinations of reliability and correlation metrics having valid solutions to equation \ref{equation:rpc_solutions}.

Taken all together, our results indicate that the answer to the question ``Are ECMWF subseasonal forecasts reliable?'' can be sensitive to the choice of atmospheric circulation index, lead time, time-averaging period, and the reforecast dataset used for verification. Our Tco319 large-ensemble reforecasts exhibit some evidence for unreliability of monthly mean wintertime NAO ($\textnormal{RPC}\approx 1.5$, $\alpha \approx 1.8$) when evaluated over the period 2001-2020. Daily mean NAO forecasts from the same dataset also show anomalously high RPC values at some subseasonal lead times, though these coincide with statistically paradoxical combinations of correlation and reliability metrics. These results are qualitatively and quantitatively reproduced in lower-resolution Tco199 100-member reforecasts for the same start dates (figure \ref{fig:tco199_100member}). However, these results do not generalise to 10-member Tco199 reforecasts with the same IFS cycle evaluated over 3120 start dates for the period 1959-2023. Given that these extended reforecasts are run with the same IFS model as the large-ensemble simulations, one might expect that systematic deficiencies, including SNP-like unreliability, would become more evident when evaluated over a much larger sample of start dates. In contrast, these extended reforecasts seem to be remarkably well-calibrated such that $\textnormal{RPC}\approx 1$ for NAO, PNA, and PVORTEX across all subseasonal lead times. 

Our findings are broadly consistent with previous studies highlighting the sensitivity of SNP diagnostics to evaluation period \citep{weisheimer2019confident}, and indicate that both observational sampling uncertainties and non-stationarity may play important roles in subseasonal reliability assessments. Based on the statistical considerations in section \ref{section:statistics} and our analysis of subseasonal reforecasts in sections \ref{section:results_uncalibrated}-\ref{section:monthly_nao}, we highlight several methodological details that we believe are important for the evaluation of reliability and signal-to-noise properties in the presence of large sampling uncertainties: 

\begin{enumerate}

      \item Evaluation of SNP-like behaviour should include careful evaluation of the climatological reliability and unbiased ensemble variance reliability conditions described in section \ref{section:reliability} and all relevant sample statistics should include uncertainty estimates. Of particular importance is the uncertainty in the observed variance ($\sigma_T^2$), which is insensitive to ensemble size and can be reduced through the use of longer reforecast periods and/or more frequent initialization \citep[e.g.][]{shi2015impact, weisheimer2019confident}. 

      \item Equation \ref{equation:rpc_solutions} provides a general theoretical framework to collectively interpret sample estimates of $r_{mo}$, SER, VR, and RPC. The presence or absence of physically admissible solutions (i.e. real-valued and non-negative variances) provides a mechanism to identify when sampling uncertainties preclude naive interpretation of sample estimates as converged population quantities. Importantly, the absence of solutions does not prove perfect reliability. It does, however, demonstrate that at least one of the sample estimates of $r_{mo}$, SER, or VR is substantially contaminated by sampling noise, such that the observed combination cannot be interpreted as converged population quantities. This diagnostic may be particularly valuable for seasonal-to-decadal forecasts where limited sample sizes make frequentist hypothesis testing challenging.
      
      \item  The optimal (affordable) balance of start dates and ensemble members should be carefully considered when designing (re)forecast datasets to evaluate ensemble reliability and signal-to-noise properties. The reforecast configuration that minimises sampling uncertainties likely depends on the intrinsic predictability of the process under investigation (e.g. figures \ref{fig:rpc_vs_ser_vr} and \ref{fig:rpc_vs_M_and_N}). In some situations, an increased number of independent start dates ($M$), which impacts both observation and model sampling uncertainties, could be more useful than increased ensemble size ($N$).

      \item Large ensembles are necessary to extract small predictable signals from noise in real-time forecasts, but they are not necessary to evaluate some aspects of statistical consistency with observations. Throughout this study, RPC, SER, and VR are calculated using unbiased estimators that account for systematic effects of both ensemble size and, in the case of anomalies, the sample size of the reference climatology \citep{leutbecher2008ensemble, roberts2025unbiased}.
      
      \item The correlation-based definition of RPC is unbiased with ensemble size provided that estimates of $r_{mo}$ and $r_{mm(k)}$ are defined such that they are exchangeable when the underlying forecast anomalies are reliable (figure \ref{fig:biased_vs_unbiased_rpc}). One way to ensure this exchangeability is to calculate $r_{mm(k)}$ between the ensemble mean and an excluded member (equation \ref{equation:model_member_correlation}) and $r_{mo}$ using an ensemble size of $N-1$ for consistency with $r_{mm(k)}$ (equation \ref{equation:correlation}). Accounting for these effects is essential for fair comparison of our 10-member and 100-member forecast configurations. 

      \item A simple approach to ensure that anomaly-based statistics are unbiased with respect to climatology sample size is to construct forecast anomalies separately for each member \citep{roberts2025unbiased}. This approach has no impact on ensemble means, but ensures that forecast member anomalies remain statistically exchangeable with observed anomalies if the underlying raw forecasts are perfectly reliable. This method of anomaly calculation does not affect estimates of $r_{mo}$ but impacts estimates of ensemble spread, total anomaly variance, $r_{mm(k)}$, and $\overline{r_{mm}}$. Nevertheless, these effects are small in the datasets considered in this study and do not materially impact our conclusions.  

      \item The correction for the signal-to-noise paradox can be interpreted as a reliability calibration and an apparent SNP can occur because the predictable signal is too weak (i.e. the diagnosed value of $\alpha > 1$) and/or the unpredictable noise is too large (i.e. the diagnosed value of $\beta < 1$). However, such methods are vulnerable to overfitting to the available observational data and should be accompanied by uncertainty estimates for the derived calibration parameters \citep[e.g.][]{siegert2016bayesian}. The most robust way to test the validity of such reforecast-derived calibration parameters is to demonstrate that they yield tangible improvements in the probabilistic skill of real-time forecasts that lie outside the training sample.

\end{enumerate}

Finally, we emphasise that correlation-based diagnostics alone provide an incomplete description of the reliability characteristics of an ensemble forecast system and many combinations of SER and VR can satisfy the constraint that RPC=1 (figure \ref{fig:rpc_vs_ser_vr}). In contrast, the fair version of the CRPS is a proper score \citep{ferro2014fair, leutbecher2021understanding} that is both unbiased with ensemble size and minimized in the limit that $\text{RPC} \to 1$ as $\text{SER},\text{VR} \to 1$ (figure \ref{fig:rpc_vs_ser_vr}a). There is thus no inconsistency between the objectives of eliminating an apparent signal-to-noise paradox and ensemble forecast development guided by unbiased evaluation of forecast reliability and optimization of proper scores. 

\section*{Acknowledgements}
Data from the ERA5 reanalysis are available to download from \url{https://www.ecmwf.int/en/forecasts/dataset/ecmwf-reanalysis-v5}. The 100-member ECMWF IFS reforecasts used in this study are available from \url{https://apps.ecmwf.int/ifs-experiments/rd/hsff/} and \url{https://apps.ecmwf.int/ifs-experiments/rd/ix2q/}. The 10-member extended IFS reforecasts used in this study are available from \url{https://apps.ecmwf.int/ifs-experiments/rd/hnll/}.

\section*{Conflict of interest}
The authors declare no conflict of interest.

\appendixpage
\begin{appendices}
\section{RPC from correlation and reliability metrics}
We start from definitions of VR and $\textnormal{SER}^2$, which correspond to the climatological and ensemble variance reliability conditions described in section \ref{section:reliability}:

\begin{equation}
    \label{equation:VR}
    \textnormal{VR} = \frac{\sigma_z^2}{\sigma_T^2}, \\
\end{equation}

\begin{equation}
\label{equation:SER}
    \textnormal{SER}^2 = \frac{\sigma_z^2 - \sigma_{\left<z\right>}^2 } {  \mathbb{E} \left[   \left( z_{T,j} - \left<z_{.,j}\right>_N \right)^2 \right] }. \\
\end{equation}
Substituting equation \ref{equation:VR} into equation \ref{equation:SER} and expanding yields a quadratic expression for $\sigma_{\left<z\right>}$

\begin{equation}
\label{equation:quadratic}
\left( \textnormal{SER}^2 + 1 \right) \sigma_{\left<z\right>}^2 - 2 \rho_{mo} \textnormal{SER}^2  \sigma_T \sigma_{\left<z\right>} + \left( \textnormal{SER}^2 - \textnormal{VR} \right) \sigma^2_T = 0, \\
\end{equation}
where $\rho_{mo}$ is the population correlation between observations and ensemble means, and we have exploited that $\mathbb{E}\left[z_{T,j}\right] = \mathbb{E}\left[\left<z_{.,j}\right>_N \right] = 0$. This equation has two solution families ($\sigma_{\left<z\right>:1}$ and $\sigma_{\left<z\right>:2}$), defined by

\begin{equation}
\label{equation:quadratic_solutions}
\sigma_{\left<z\right>:1,2} = \frac{  \rho_{mo} \sigma_T \textnormal{SER}^2  \pm \sigma_T \sqrt{\Delta} }{ 1 + \textnormal{SER}^2 },\\
\end{equation}
where 

\begin{equation}
\label{equation:discriminant}
\Delta \equiv (\rho_{mo}^2 -1) \textnormal{SER}^4 + (\textnormal{VR} - 1) \textnormal{SER}^2 + \textnormal{VR}.
\end{equation}
Real-valued solutions for $\sigma_{\left<z\right>}$ require that the discriminant is non-negative such that

\begin{equation}
\label{equation:real_value_condition}
\textnormal{VR} \ge \frac{\textnormal{SER}^2 + (1 - \rho_{mo}^2)\textnormal{SER}^4 } {1 + \textnormal{SER}^2}.
\end{equation}
Furthermore, physically admissible solutions (i.e. non-negative variances) are obtained only under the additional conditions:

\[
\rho_{mo} \ge 0\;\;
\begin{cases}
\sigma_{\left<z\right>:1} \ge 0 \ & \text{always}, \\[6pt]
\sigma_{\left<z\right>:2} \ge 0 &\text{if } \textnormal{SER}^2 \ge \textnormal{VR},
\end{cases}
\]

\[
\rho_{mo} < 0\;\;
\begin{cases}
\sigma_{\left<z\right>:1} \ge 0 &\text{if } \textnormal{VR} \ge \textnormal{SER}^2,\\[6pt]
\sigma_{\left<z\right>:2} \ge 0 &\text{never}.
\end{cases}
\]
From these solutions to $\sigma_{\left<z\right>}$, one can then derive solutions for population model-model correlations ($\rho_{mm}$) that are independent of $\sigma_t$

\begin{align}
\label{equation:r_mm_solutions}
\rho_{mm:1,2} = \frac{ \rho_{mo} \textnormal{SER}^2  \pm  \sqrt{\Delta} }{ \sqrt{\textnormal{VR}} \left( 1 + \textnormal{SER}^2 \right)},
\end{align}
Substitution of $\rho_{mm:1,2}$ into equation \ref{equation:rpc_corr} then yields the two families of RPC solutions given by equation \ref{equation:rpc_solutions}.

\end{appendices}

\newpage
\bibliographystyle{rss}
\bibliography{references}

\newpage

\newpage

\renewcommand{\thefigure}{S1}
\begin{figure}[htbp]
    \centering
    \begin{minipage}{0.48\textwidth}
        \centering
        \includegraphics[width=\textwidth]{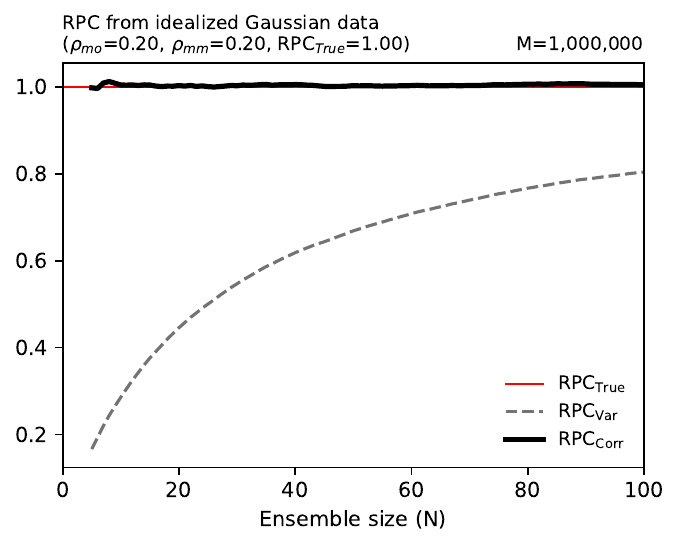}
    \end{minipage}
    \hfill
    \begin{minipage}{0.48\textwidth}
        \centering
        \includegraphics[width=\textwidth]{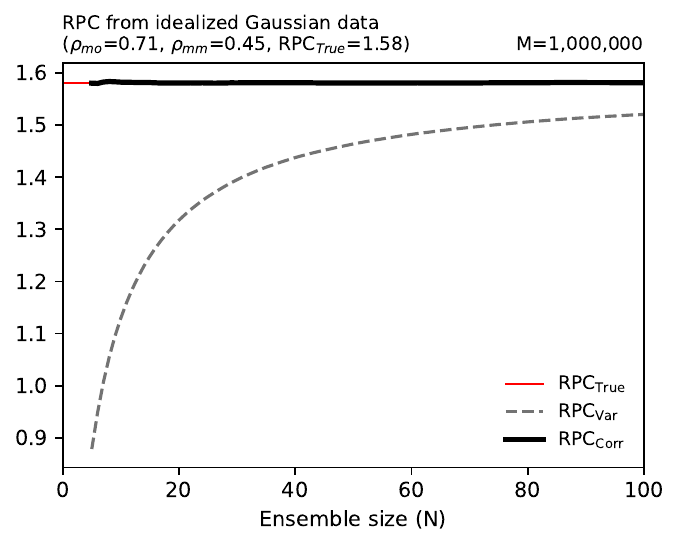}
    \end{minipage}
    \caption{A comparison of biased and unbiased RPC estimators as a function of ensemble size ($N$) applied to idealised Gaussian forecast data with $M = 1{,}000{,}000$ independent cases. $\textnormal{RPC}_{\textnormal{Var}}$ is calculated using equation 8 in the main text and is biased low for finite $N$. In contrast, $\textnormal{RPC}_{\textnormal{Corr}}$ is calculated using equation 9 from the main text. When model-observation and model-model correlations are evaluated using equations 5 and 6, respectively, this definition of $\textnormal{RPC}_{\textnormal{Corr}}$ is unbiased with ensemble size for both reliable (left) and unreliable (right) forecast scenarios.}
    \label{fig:biased_vs_unbiased_rpc}
\end{figure}

\renewcommand{\thefigure}{S2}
\begin{figure}[!htbp]
    \
    \includegraphics[width=15cm]{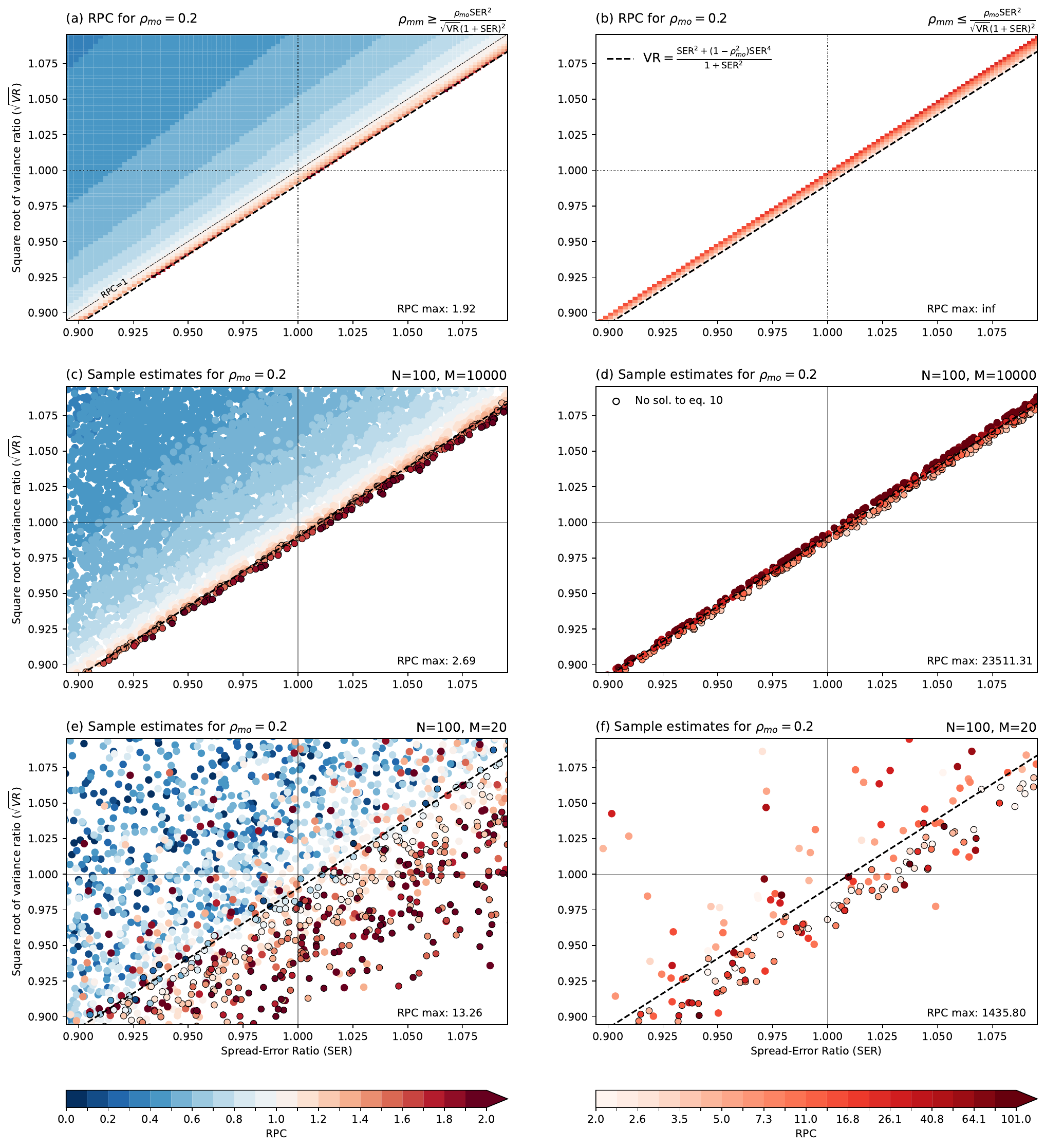}
    \center
    \caption{As figure 1, but for $\rho_{mo}=0.2$.}
    \label{fig:rpc_vs_ser_vr_rho02}
\end{figure}

\renewcommand{\thefigure}{S3}
\begin{figure}[!htbp]
    \
    \includegraphics[width=15cm]{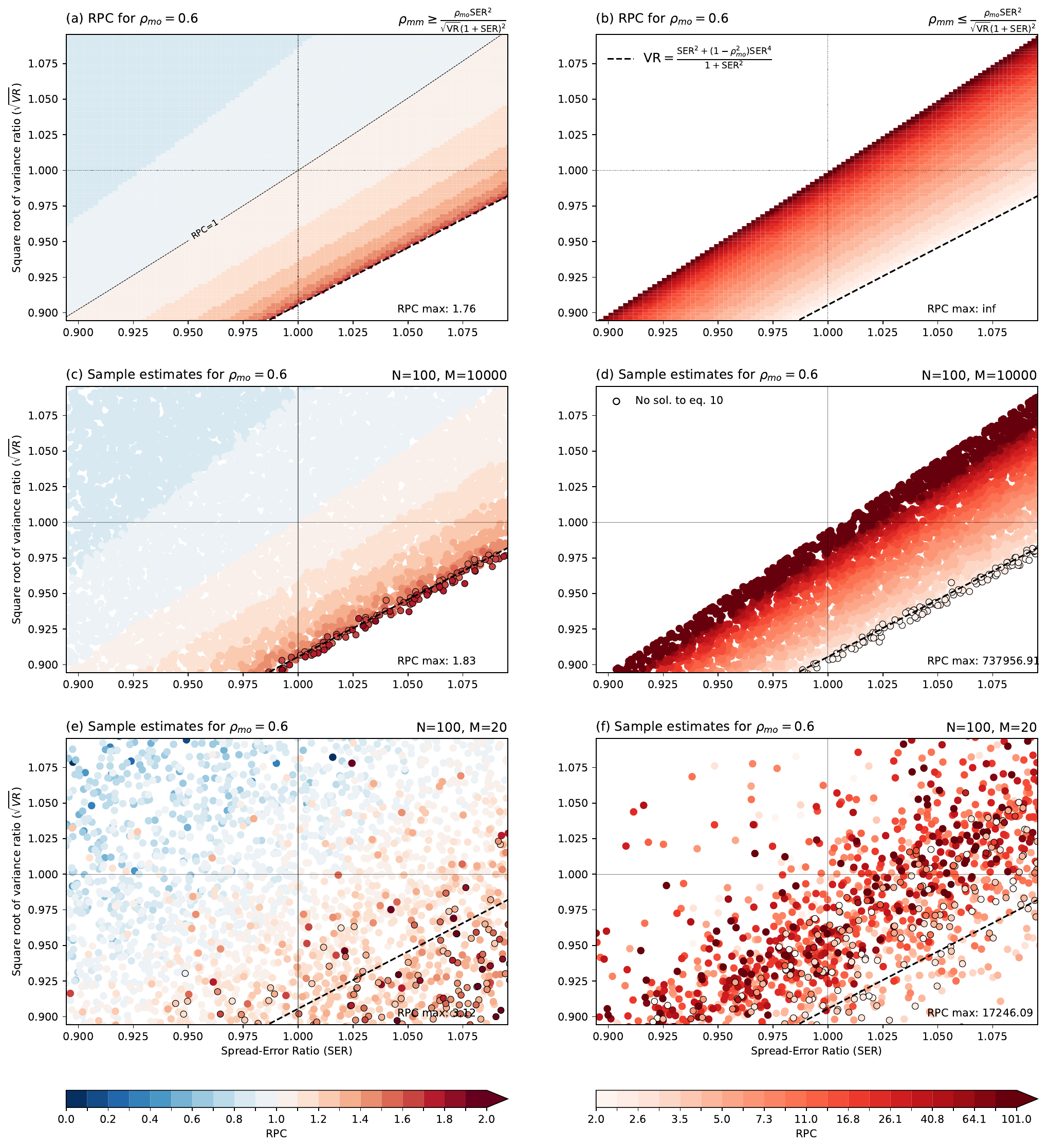}
    \center
    \caption{As figure 1, but for $\rho_{mo}=0.6$.}
    \label{fig:rpc_vs_ser_vr_rho06}
\end{figure}

\renewcommand{\thefigure}{S4}
\begin{figure}[!htbp]
    \includegraphics[width=15cm]{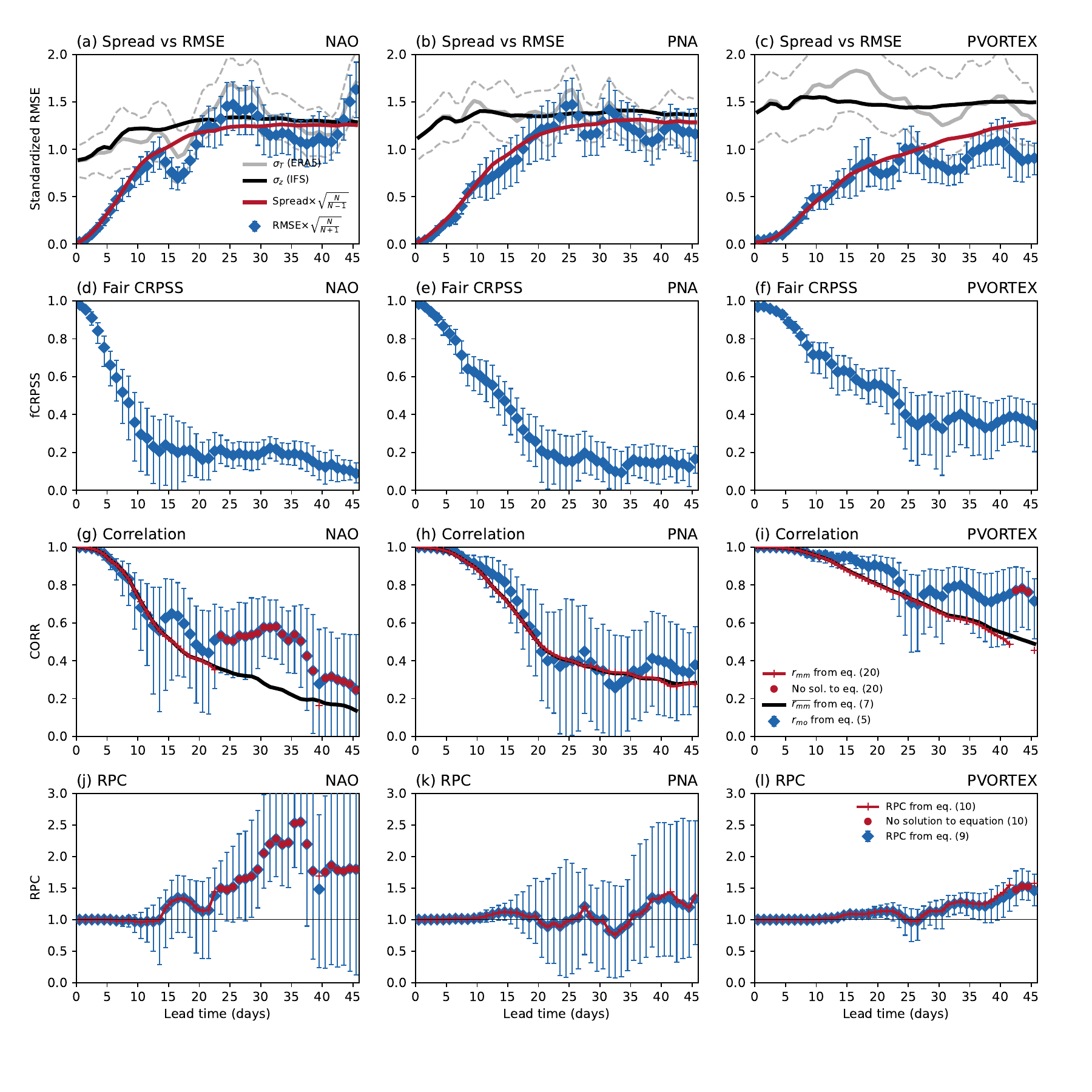}
    \center 
    \caption{As figure 3, but for Tco319 reforecasts covering the period 2001-2020 initialized during February and November such that $M=40$ and $N=100$}.
    \label{fig:tco319_feb_nov}
\end{figure}

\renewcommand{\thefigure}{S5}
\begin{figure}[!htbp]
    \includegraphics[width=15cm]{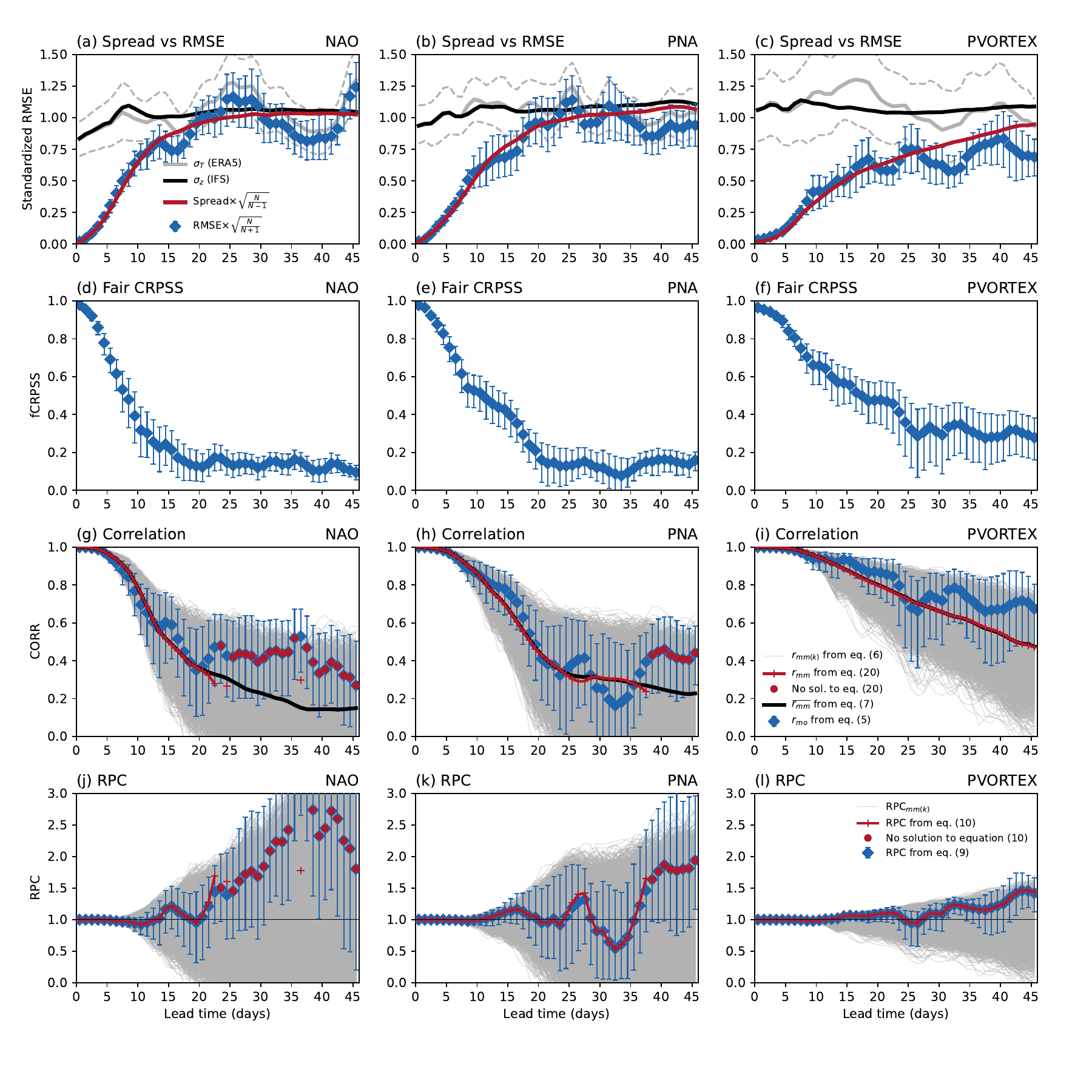}
    \center 
    \caption{As figure 3, but for Tco199 reforecasts covering the period 2001-2020 with $M=80$ and $N=100$}.
    \label{fig:tco199_100member}
\end{figure}

\renewcommand{\thefigure}{S6}
\begin{figure}[!htbp]
    \includegraphics[width=15cm]{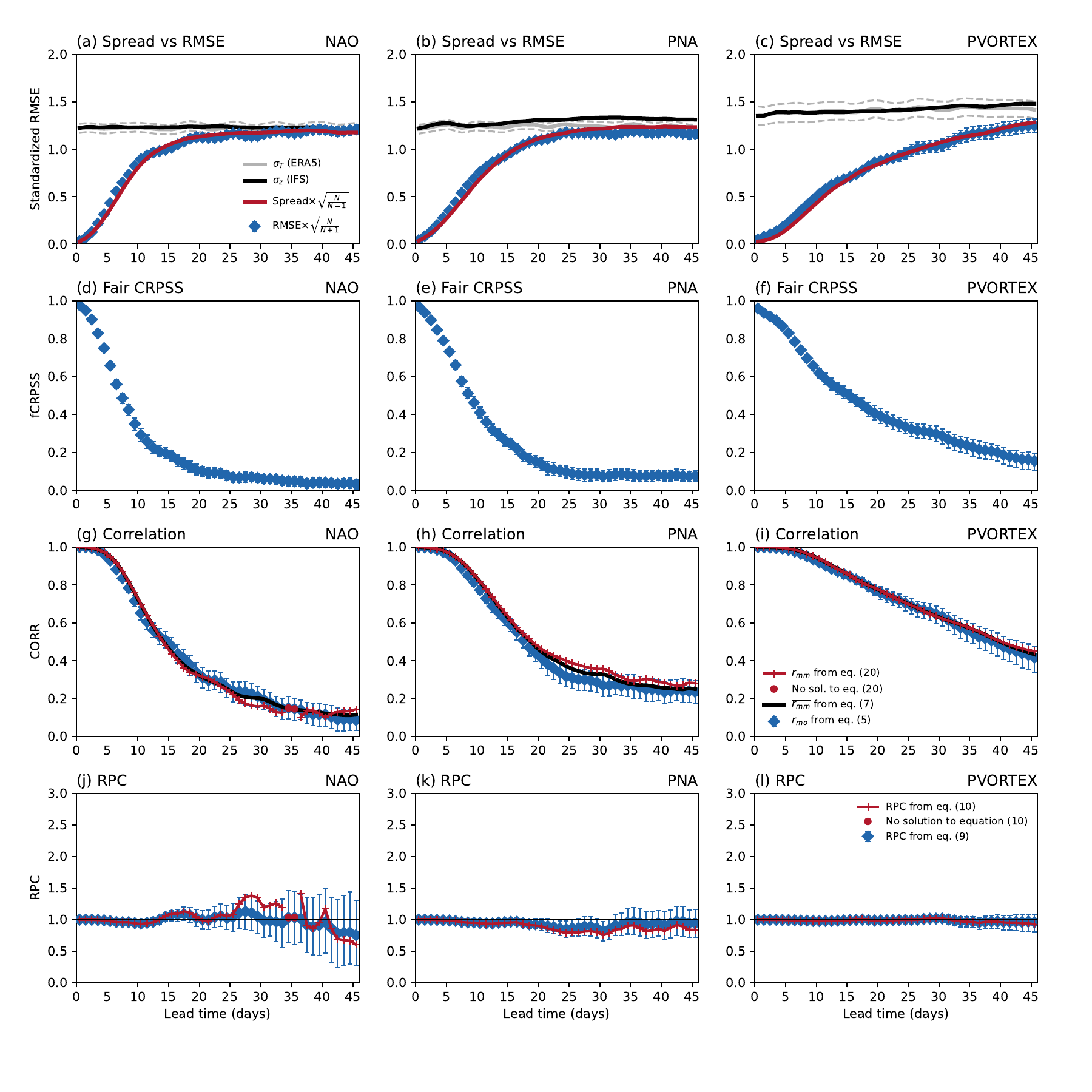}
    \center 
    \caption{As figure 5, but for Tco199 reforecasts covering the period 1959-2023 initialized during the October-March extended winter period such that $M=1560$ and $N=10$.}
    \label{fig:tco199_extended_winter}
\end{figure}

\renewcommand{\thefigure}{S7}
\begin{figure}[!htbp]
    \includegraphics[width=15cm]{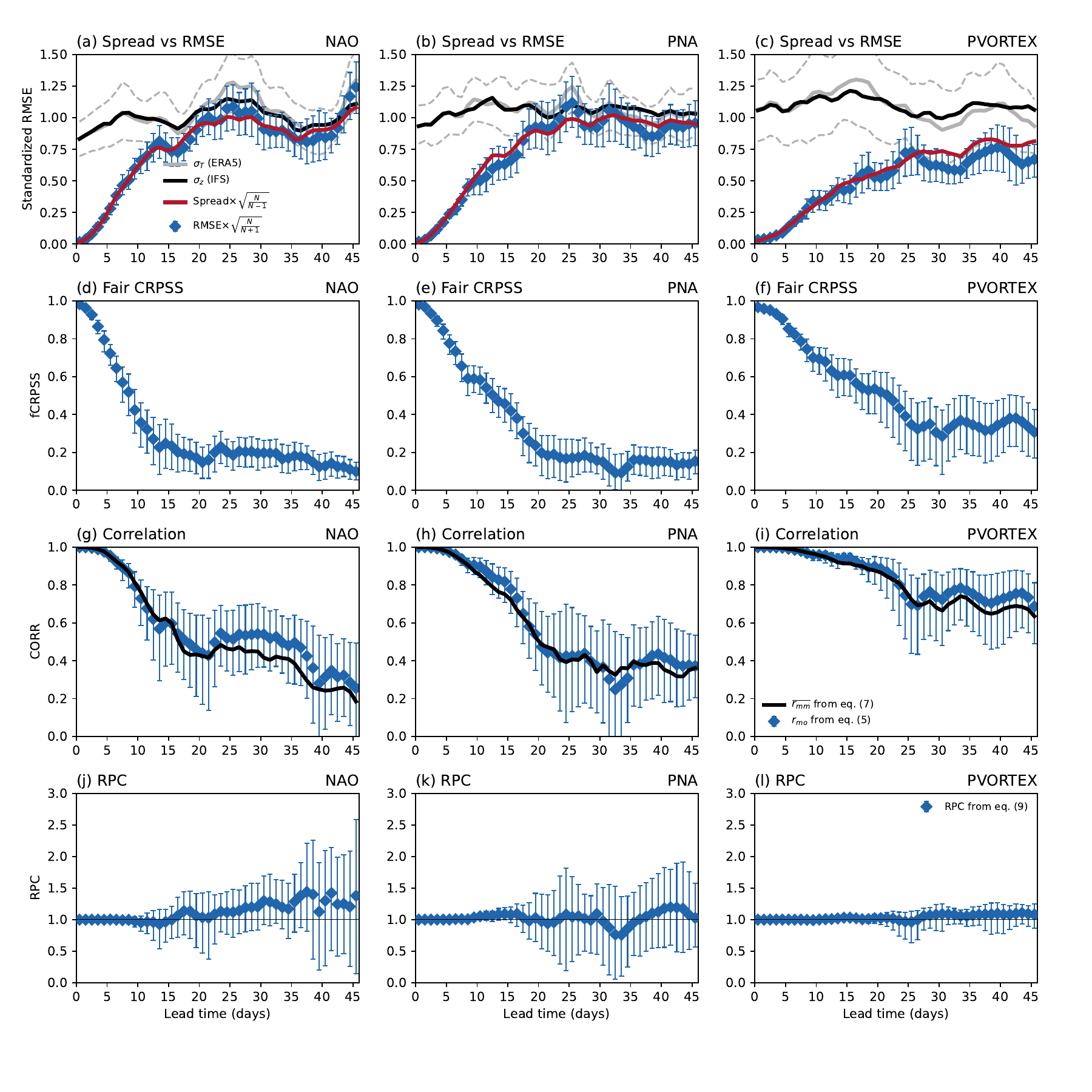}
    \center 
    \caption{As figure 3, but for circulation indices derived from calibrated grid-point anomalies.}
    \label{fig:calibrated_gridpoint}
\end{figure}

\end{document}